\documentclass[twocolumn]{aastex701}

\usepackage{appendix}

\usepackage{amsmath}
\usepackage{comment}
\accepted{August 26, 2026}

\def\gtrsim{\lower 2pt \hbox{$\, \buildrel {\scriptstyle >}\over
{\scriptstyle \sim}\,$}}
\def\lesssim{\lower 2pt \hbox{$\, \buildrel {\scriptstyle <}\over
{\scriptstyle \sim}\,$}}

\shorttitle{}

\graphicspath{{./}{figures/}}

\begin{document}
\title{Energy Partition in AGN-driven Bubbles of NGC 4438: From Nuclear Bubbles to a Galaxy-scale Outflow}

\correspondingauthor{Luan Luan}
\author[0009-0008-2940-6166]{Luan Luan}
\email[show]{luanluan@pmo.ac.cn}
\affiliation{Purple Mountain Observatory, Chinese Academy of Sciences, 10 Yuanhua Road, Nanjing 210023, China}

\author[0000-0001-6239-3821]{Jiang-Tao Li}
\email[show]{pandataotao@gmail.com}
\affiliation{Purple Mountain Observatory, Chinese Academy of Sciences, 10 Yuanhua Road, Nanjing 210023, China}

\author[0009-0006-3887-8988]{Jianghui Xu}
\email{jianghuixu@mail.ustc.edu.cn}
\affiliation{Department of Astronomy, University of Science and Technology of China, Hefei, Anhui 230026, China}
\affiliation{School of Astronomy and Space Science, University of Science and Technology of China, Hefei 230026, China}
\affiliation{Hamburger Sternwarte, Universit\"at Hamburg, Gojenbergsweg 112, D-21029 Hamburg, Germany}

\author[0000-0001-7254-219X]{Yang Yang}
\email{yangyang.astro@gmail.com}
\affiliation{Xiangtan University, Xiangtan 411105, Hunan, China}

\author[0000-0003-4286-5187, gname=Guilin, sname=Liu]{Guilin Liu}
\affiliation{Department of Astronomy, University of Science and Technology of China, Hefei, Anhui 230026, China}
\affiliation{School of Astronomy and Space Science, University of Science and Technology of China, Hefei 230026, China}
\email{glliu@ustc.edu.cn}

\author[0000-0003-1474-8899]{Fulai Guo}
\affiliation{Shanghai Astronomical Observatory, Chinese Academy of Sciences, 80 Nandan Road, Shanghai 200030, China}
\email{fulai@shao.ac.cn}

\author[0000-0002-9279-4041, gname='Q. Daniel', sname=Wang]{Q. Daniel Wang}
\affiliation{Department of Astronomy, University of Massachusetts, Amherst, MA 01003, USA}
\email{wqd@umass.edu}

\begin{abstract}
Jets launched by accreting supermassive black holes represent a major mode of active galactic nucleus (AGN) feedback. However, how their energy is divided among bulk kinetic motion, thermal gas, magnetic fields, cosmic rays (CRs), and radiation -- and how this distribution changes with spatial scale -- remains poorly constrained.
NGC~4438 provides a unique laboratory for probing this evolution, hosting two 200-pc-scale nuclear bubbles and a lopsided $\sim10$~kpc galaxy-scale outflow plausibly associated with the same AGN.
We present a multi-wavelength analysis to investigate the morphology, radiation mechanisms, and energetics of these structures.
Joint radio–X-ray modeling shows that the non-thermal emission in the nuclear bubbles may require two distinct populations of cosmic-ray electrons, suggesting that in addition to shock acceleration at the bubble rim, the highest-energy particles may be linked to acceleration processes closer to the unresolved central engine.
A spatially resolved energy inventory reveals that bulk kinetic energy dominates the current energy budget of the nuclear bubbles, while roughly half of the injected energy has already been transformed into thermal, CR, and magnetic energy, as well as radiative losses.
Across all bubble sizes, the thermal and magnetic pressures are consistent within the uncertainties, implying that magnetic fields remain dynamically significant on all examined spatial scales.
Furthermore, the empirical correlation between radio luminosity and jet power, established for kiloparsec-scale jet bubbles \citep{MerloniHeinz2007}, matches the energetics of the galaxy-scale outflow but substantially overestimates the power of the 200-pc-scale nuclear bubbles, underscoring the scale dependence of jet energy dissipation.
\end{abstract}

\keywords{\uat{active galactic nuclei}{16}, \uat{galaxy jets}{601}, \uat{galaxy winds}{626}, \uat{interstellar magnetic fields}{845}, \uat{interstellar synchrotron emission}{856}, \uat{plasma astrophysics}{1261}}

\section{Introduction}
Relativistic jets from accreting supermassive black holes are widely invoked as a primary channel of mechanical AGN feedback, capable of heating gaseous halos, regulating cooling, and driving multi-phase outflows across a broad range of galaxy masses and environments \citep[e.g.,][]{McNamaraNulsen2007,Fabian2012,HeckmanBest2014}.
While the global impact of jet feedback is well established in massive ellipticals and clusters through X-ray cavities and shocks \citep[e.g.,][]{Boehringer1993,Birzan2004,Rafferty2006}, key physical questions remain unresolved: how the injected jet energy is partitioned into (i) bulk kinetic energy of outflowing gas, (ii) thermal energy of shock-heated plasma, (iii) magnetic and cosmic-ray energy, and (iv) radiative losses, and how these components evolve with distance from the nucleus and ultimately drive large-scale outflows \citep[e.g.,][]{BegelmanCioffi1989,Churazov2002,Croston2005,HardcastleCroston2020}.
These energy transformations are mediated by shocks, turbulent dissipation, and mixing between the jet plasma and the surrounding ISM/CGM; however, the relative importance and detailed interplay of these processes remain uncertain, particularly for low-power jets propagating through dense, multi-phase disk environments \citep[e.g.,][]{WagnerBicknellUmemura2012,Morganti2017,Cielo2018}. 

A major observational challenge is that different wavebands trace different phases and energy reservoirs of the outflow. X-ray spectroscopy constrains the thermodynamic state and energy content of hot gas, radio synchrotron emission traces relativistic electrons and magnetic fields, optical recombination lines trace warm ionized gas and radiative shocks, and molecular lines trace the cold phase as well as its bulk and turbulent motions \citep[e.g.,][]{Veilleux2005,RupkeVeilleuxSanders2005,Martin2005}. 
Another challenge is that structures observed on different spatial scales may trace different stages of jet-driven feedback. 
Compact circumnuclear bubbles on sub-kiloparsec scales likely represent an early phase of jet interaction with the dense nuclear ISM, and may therefore depend sensitively on the local gas conditions and the instantaneous energy injection rate. 
Such structures appear to be relatively rare, presumably because this phase is short-lived. 
By contrast, larger bubbles and outflows on kiloparsec to tens-of-kiloparsec scales more often reflect the cumulative impact of sustained energy injection into the galactic disk, halo, or circumgalactic medium \citep[e.g.,][]{Fabian2012,McNamaraNulsen2007,Li2008,Heald2022,Li2019,Li2024}. 
It therefore remains unclear how jet-driven bubbles evolve from the compact circumnuclear phase, in which they primarily interact and mix with the dense ISM, to the large-scale outflow phase, in which they increasingly couple to the galactic halo or circumgalactic medium. 
This dynamical evolution is fundamentally governed by the progressive redistribution of injected energy among kinetic, thermal, magnetic, CR and radiative components.

\begin{figure*}[t]
    \centering
    \includegraphics[width=\textwidth]{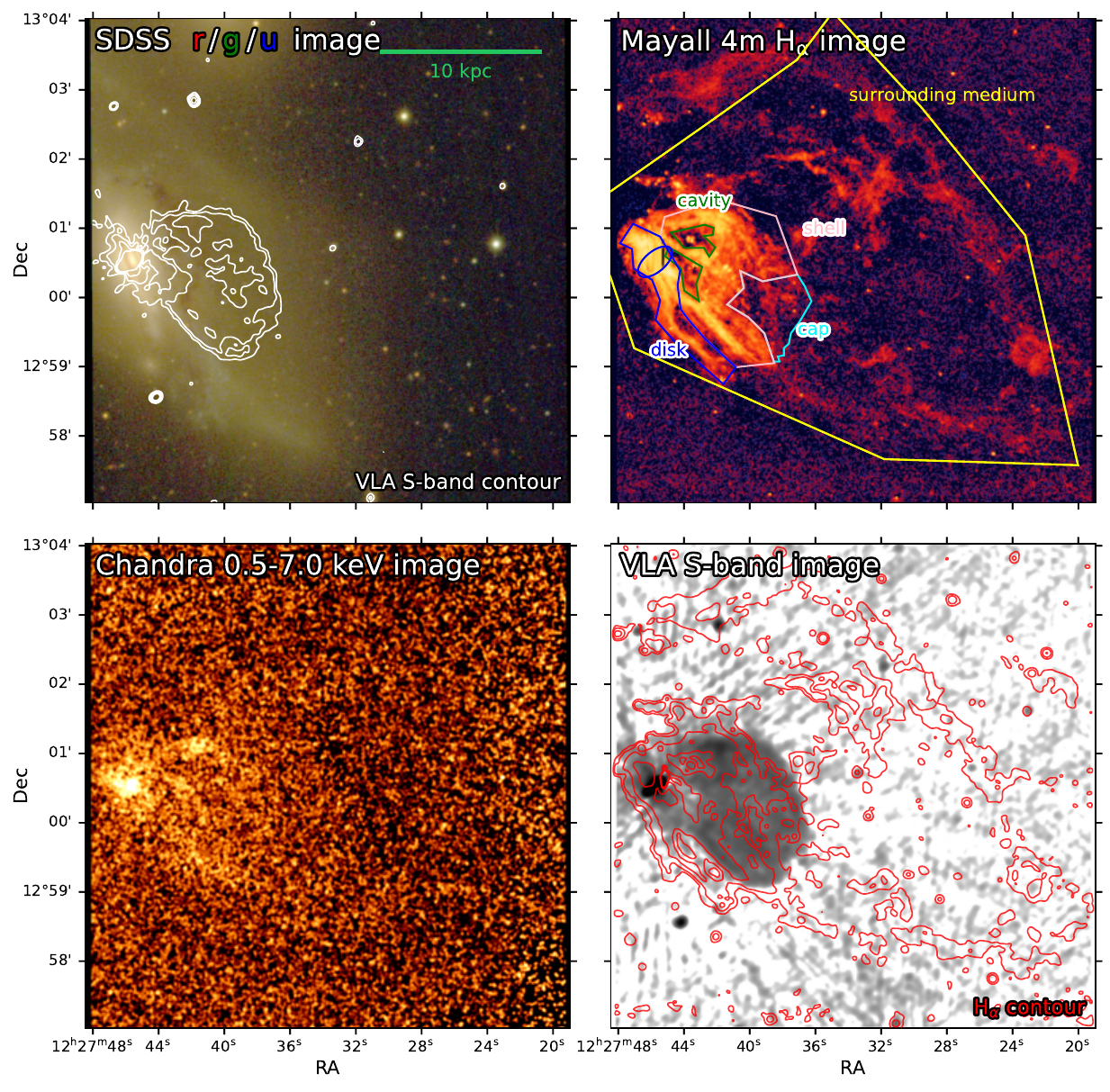}
\caption{
Multi-wavelength view of the galaxy-scale outflow in NGC~4438.
The top-left panel shows an SDSS $r/g/u$ three-color composite image, with white contours tracing the VLA $S$-band radio emission.
The top-right panel shows the deep Mayall 4-m H$\alpha$+[N\,II] image from \citet{Kenney2008}. The colored outlines mark the regions used for spectral extraction, including the cavity, shell, and cap associated with the galaxy-scale outflow, together with the disk and surrounding-medium regions. The central blue ellipse marks the region excluded to avoid contamination from the AGN and the nuclear bubbles.
The bottom-left panel shows the \textit{Chandra} 0.5--7.0~keV image, smoothed for display.
The bottom-right panel shows the VLA $S$-band radio continuum image, with red contours tracing the H$\alpha$+[N\,II] emission.
}
\label{fig:outflow}
\end{figure*}

NGC~4438, a disturbed spiral galaxy in the M86 subgroup of the Virgo Cluster at a distance of 14.4 Mpc \citep[1 arcsec $\simeq$ 70 pc, ][]{Ekholm2000}, provides a rare opportunity to address this problem observationally, as it hosts both compact nuclear bubbles on sub-kiloparsec scales and a much larger 10-kpc scale outflow structure. Given the relatively low present-day star formation rate of the galaxy and the absence of evidence for a recent powerful starburst \citep[e.g.,][]{Vargas2019}, AGN activity is widely considered the most plausible driver of these structures, likely associated with the same central engine. 
On sub-kiloparsec scales, \citet{Li2022} identify and characterize a compact X-ray bubble in the nuclear region, measuring the temperature, density, luminosity, and energetic of the hot plasma and arguing for an AGN origin. 
They also reported diffuse hard X-ray emission associated with the bubble and suggested that it may arise from synchrotron radiation of extremely high-energy (TeV) electrons. Complementary radio studies by \citet{PuigSubira2026}, using VLA and e-MERLIN continuum observations, revealed detailed jet morphology and spectral index structure and provided estimates of the jet energetics and power associated with the nuclear outflow. 
On larger scales, NGC~4438 also hosts a prominent, 10-kpc scale outflow/bubble structure visible in X-rays and optical line emission, plausibly representing the downstream outcome of sustained mechanical energy injection into the surrounding ISM and halo (see Figure \ref{fig:outflow} and \citet{Machacek2004}).
The coexistence of compact nuclear bubbles and a galaxy-scale outflow in a single system enables a controlled, multi-scale test: by treating the nuclear bubbles as snapshots of the current energy injection and early-stage coupling, and the galaxy-scale outflow as the accumulated response of the galaxy to longer-term driving, we can directly constrain how the energy partition changes with scale.

These previous works motivated a joint analysis of the nuclear bubbles and the galaxy-scale outflow in NGC~4438, which may provide a more complete view of how AGN-driven energy is distributed among thermal, magnetic, cosmic-ray, and radiative components across different feedback structures within the same system. 
In particular, joint radio–X-ray analysis of the nuclear bubbles can help constrain the non-thermal electron population and the associated particle acceleration processes \citep{Li2019}. 
Comparing the nuclear bubbles with the galaxy-scale outflow further provides an empirical view of how injected energy is redistributed across spatial scales.

In this paper, we combine deep \textit{Chandra} observations with multi-scale H$\alpha$ imaging, including \textit{HST} data for the nuclear bubbles and ground-based narrowband imaging from the KPNO 4 m telescope for the galaxy-scale outflow, together with VLA radio continuum data in both the A and C configurations, in order to investigate jet-driven feedback in NGC~4438 across a wide range of spatial scales.
Section~\ref{sec:data} describes the reduction of the X-ray and radio data used in this work. 
In Section~\ref{sec:analysis} we present the analysis of the jet-driven structures, including the multi-wavelength morphology of the nuclear bubbles, the joint radio–X-ray spectral modeling of their non-thermal emission, and the characterization of the galaxy-scale outflow.
In Section~\ref{sec:discussion} we investigate the energetics of the system by deriving the resolved energy budget of the nuclear bubbles and comparing it with the energy content and dynamical properties of the kiloparsec-scale outflow. 
Our main conclusions are summarized in Section~\ref{sec:conclusion}. 

\section{Data Reduction} \label{sec:data}
\subsection{Chandra X-ray data reduction}

This study employs the same set of archival \emph{Chandra}/ACIS observations of NGC~4438 as compiled in Table~1 of \citet{Li2022}, yielding a total effective exposure of 124.93~ks. 
The datasets were reprocessed using procedures analogous to those in \citet{Li2022}, including standard calibration, reprojection, and co-addition of the individual observations. 
Because our analysis jointly considers the nuclear bubbles and the surrounding galaxy-scale outflow, we focus below only on those elements of the X-ray analysis that deviate from, or are especially pertinent compared to, the earlier work.

\begin{figure*}[!htbp]
\centering
\includegraphics[width=\textwidth]{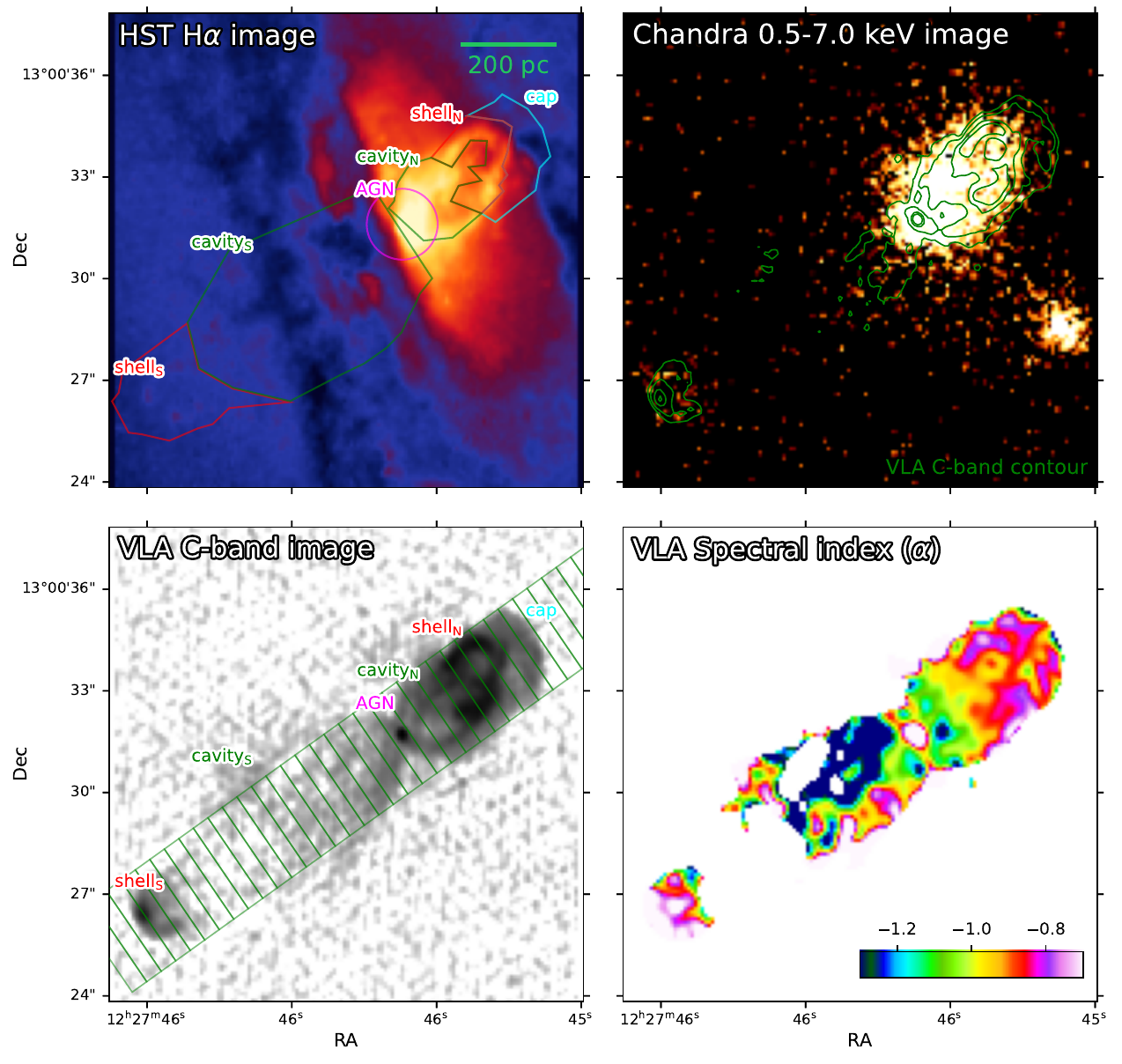}
\caption{
Multi-wavelength view of the two nuclear bubbles in the central region of NGC~4438.
The top-left panel shows the \textit{HST} H$\alpha$ image. The colored outlines mark the regions used for spatially resolved spectral extraction, including the shell (red), cavity (green), and cap (cyan), while the magenta circle marks the central AGN region excluded from the analysis.
The top-right panel shows the \textit{Chandra} broad-band X-ray image, with radio continuum contours overlaid to illustrate the correspondence between the X-ray and radio structures.
The bottom-left panel shows the VLA radio continuum image. The overlaid rectangular strip and its subdivisions indicate the regions used to construct the one-dimensional profiles presented in Figure~\ref{fig:majoraxis_profile}.
The bottom-right panel shows the radio spectral-index map, with radio continuum contours overlaid to indicate the morphology of the nuclear bubbles.}
\label{fig:north_regions}
\end{figure*}

To isolate the diffuse X-ray emission associated with the kpc-scale outflow, discrete X-ray point sources were identified with the wavelet-based algorithm \texttt{wavdetect} on exposure-corrected images in the broad, soft, and hard bands, using wavelet scales of 1, 2, 4, and 8 pixels and a false-positive probability threshold of $10^{-6}$. 
Source lists from different bands were merged into a master catalog, with overlapping detections treated as a single source. 
For the large-scale outflow ($\sim2\arcmin$), all detected point sources were excluded using circular masks based on the merged \texttt{wavdetect} source extents and enlarged where necessary to account for the local point-spread function. 
In contrast, no point-source excision was applied in the central few arcseconds for the nuclear jet analysis, where such masking would remove a substantial fraction of the jet-related emission; this different treatment reflects the distinct spatial scales and scientific goals of the nuclear and extended analyses.

Spectra were extracted from the regions of interest using standard CIAO tools and combined for multiple observations when appropriate. 
The total background was treated as the sum of instrumental and astrophysical components, following the strategy described by \citet{Wang2021M83}. 
The noncosmic instrumental background was estimated using the \texttt{mkacispback} package \citep{Suzuki2021}, which constructs a model background based on ACIS stowed data and the Chandra Deep Field-South survey. 
Instrumental background spectra were generated for individual observations and combined using exposure-time weighting. 
The astrophysical (cosmic) X-ray background was estimated locally from source-free regions at large angular separations from the galaxy, ensuring negligible contamination from the extended emission of NGC~4438.

To ensure sufficient statistical quality for spectral fitting while preserving spatial information, the extracted spectra were adaptively grouped so that each spectral bin has a signal-to-noise ratio $>2$ after background subtraction. 
All spectral fitting was performed using XSPEC, and uncertainties are quoted at the $1\sigma$ confidence level unless stated otherwise.

In addition to the standard spectral models available in XSPEC, we employed a custom models \texttt{mxabs}, a multiplicative absorption model that assumes the X-ray-emitting plasma is macroscopically mixed with the absorbing cool gas, rather than being subject to purely foreground absorption. 
This model provides a more physical description of the emission--absorption geometry in galactic environments and is described in detail by \citet{LuanWang2025}.

\subsection{Radio data}\label{subsec:radio data}
To investigate the multi-scale structures of NGC~4438, we utilized VLA $S$-band (A- and C-configurations) and $C$-band (A-configuration) observations. 
The A-configuration data ($S$- and $C$-bands) were obtained under project 22A-180 (PI: J.-T. Li), while the C-configuration $S$-band data were part of project 21A-033 (PI: Y. Stein) from the Continuum Halos in Nearby Galaxies--an EVLA Survey \citep[CHANG-ES;][]{Irwin12,Irwin12b}. 
The total on-source integration times are 3.29~hr (A-$S$), 1.18~hr (A-$C$), and 3.12~hr (C-$S$).

Data reduction was performed using the Common Astronomy Software Applications package \citep[CASA, version 6.6.4;][]{CASA22}. 
Following calibration procedures similar to those described in \citet{Xu26}, we applied three rounds of phase-only self-calibration to the C-configuration $S$-band data and one round to the A-configuration $S$-band data. 
The final Stokes $I$ images were reconstructed using the \texttt{tclean} task with the \texttt{mtmfs} deconvolver (\texttt{nterms} = 2, Briggs weighting with robust=0). 
At 3.0~GHz, the $S$-band maps achieve synthesized beams of $0\farcs51 \times 0\farcs44$ (rms $5.0\,\mu{\rm Jy\,beam}^{-1}$) and $5\farcs93 \times 4\farcs87$ (rms $5.9\,\mu{\rm Jy\,beam}^{-1}$) for the A- and C-configurations, respectively. 
The 6.0~GHz $C$-band map achieves a beam of $0\farcs26 \times 0\farcs23$ (rms $4.3\,\mu{\rm Jy\,beam}^{-1}$).

For the A-configuration $S$-band polarization imaging, we applied an additional $1\arcsec$ \textit{uv}-taper. 
The resulting $Q$ and $U$ maps achieve a synthesized beam of $1\farcs02 \times 0\farcs98$ with a mean rms noise level of $6.0\,\mu{\rm Jy\,beam}^{-1}$. 
Linear polarization intensity and angle maps were derived from the Stokes $Q$ and $U$ maps, with the Ricean bias corrected following \citet{Wardle74}.

For joint radio and X-ray spectral fitting (Section \ref{sec:spectral} and \ref{sec:outflow}), we imaged the data in narrower frequency bins using \texttt{tclean} with \texttt{nterms} = 1. 
The $S$-band data were imaged per individual spectral window (spw; 128~MHz width).
All narrow-band images were corrected for the primary beam response.

Spectral index maps between the $S$- and $C$-bands were derived after smoothing all the $C$-band images (including the C-configuration data from \citealt{Li2022}) to match the corresponding $S$-band resolution. 
We estimated the equipartition magnetic field strength ($B_{\mathrm{eq}}$) following \citet{Beck05}. 
We assumed a cosmic-ray proton-to-electron ratio of $K_0 = 100$. 
To isolate the non-thermal intensity ($I_{\mathrm{nt}}$), we separated the radio continuum into thermal and non-thermal components assuming fixed spectral indices of $-0.1$ \citep{Condon92} and $-1.0$ \citep{Hota2007}, respectively (assuming 100\% non-thermal emission where the total index is steeper than $-1.0$). 
Adopting line-of-sight path lengths of $l = 8.38$~kpc for the kpc-scale and 0.24~kpc for the sub-kpc scale bubbles, the typical propagated relative error in $B_{\mathrm{eq}}$ is approximately 14\% (assuming 50\% uncertainty for $l$ and $K_0$).

\section{Analysis}\label{sec:analysis}
\begin{figure*}
    \centering
    \includegraphics[width=\textwidth]{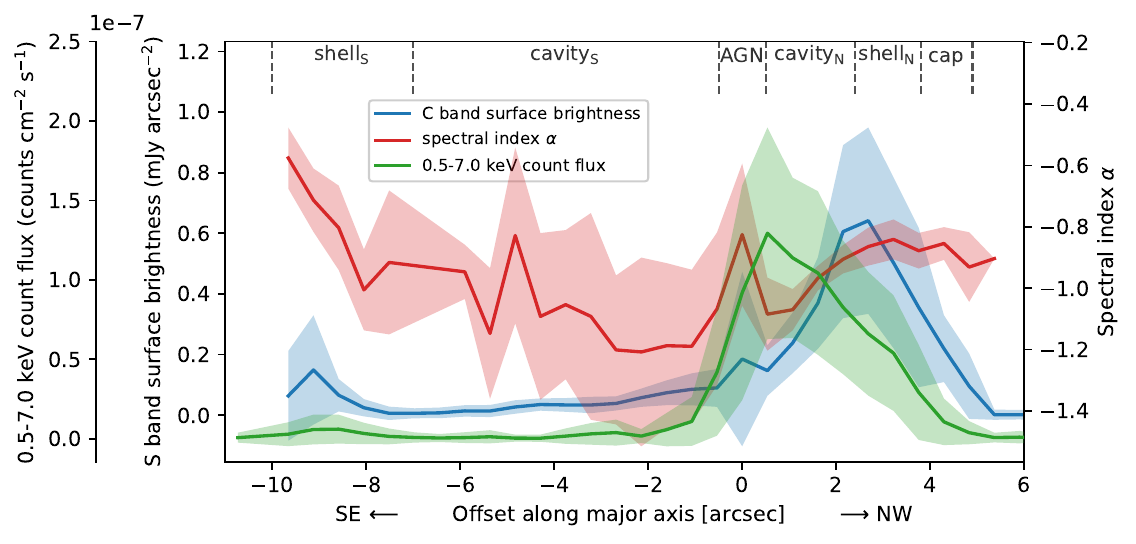}
\caption{
One-dimensional profiles measured along the nuclear bubble shown in Figure~\ref{fig:north_regions}c. The blue, red, and green curves show the $C$-band surface brightness, radio spectral index, and \textit{Chandra} broad-band (0.5--7.0 keV) count flux, respectively. At each position along the major axis, the plotted quantities are averaged over a 200 pc-wide strip in the minor-axis direction, and the error bars represent the dispersion within that interval. The labels at the top indicate the approximate locations of the shell, cavity, AGN, and cap regions defined from the two-dimensional morphology shown in Figure~\ref{fig:north_regions}a. Because these regions are irregular in shape, their extents along the profile should be regarded as approximate.
}
    \label{fig:majoraxis_profile}
\end{figure*}

\begin{deluxetable*}{lccccccc}
\tabletypesize{\scriptsize}
\tablecaption{Spectral fitting parameters and derived properties of the nuclear bubbles.\label{tab:fitting_small_bubble}}
\tablewidth{0pt}
\tablehead{
\colhead{Fitting Parameters} &
\multicolumn{3}{c}{SE nuclear bubble} &
\multicolumn{4}{c}{NW nuclear bubble}\\
\cline{2-4}\cline{5-8}
&
\colhead{whole} & \colhead{shell$_S$} & \colhead{cavity$_S$} &
\colhead{whole} & \colhead{cavity$_N$} & \colhead{shell$_N$} & \colhead{cap}
}
\startdata
area ($arcsec^2$) &46.19 &7.85 &36.79 &10.76 &3.57 &3.39 &3.80 \\
$N_{\rm H,MW}$ ($10^{22}\,\mathrm{cm^{-2}}$)
& \multicolumn{7}{c}{0.0213 (fixed)} \\
$N_{\rm H,int}$ ($10^{22}\,\mathrm{cm^{-2}}$)
& $0.34^{+0.10}_{-0.08}$ &0.34 (fixed)&0.34 (fixed)
&$0.25^{+0.09}_{-0.08}$&$0.26^{+0.14}_{-0.12}$&$0.29^{+0.16}_{-0.14}$&$0.45^{+0.32}_{-0.29}$ \\
$kT$ (keV)
& $0.73^{+0.07}_{-0.08}$ &0.73 (fixed)&0.73 (fixed)
&$0.79^{+0.03}_{-0.03}$&$0.81^{+0.04}_{-0.04}$&$0.80^{+0.05}_{-0.05}$&$0.79$(fixed) \\
$Z$ ($Z_\odot$)
& \multicolumn{7}{c}{1.0 (fixed)} \\
$\mathrm{norm}_{\rm apec}$ ($10^{-5}$)
&$0.90^{+0.46}_{-0.25}$&$0.11^{+0.04}_{-0.05}$&$0.27^{+0.08}_{-0.09}$
&$3.12^{+0.38}_{-0.31}$&$1.59^{+0.30}_{-0.22}$&$1.11^{+0.22}_{-0.17}$&$0.16^{+0.05}_{-0.05}$ \\
$\alpha_{\rm radio}$
&-&-&-&$1.113^{+0.07}_{-0.06}$&$1.23^{+0.07}_{-0.07}$&$0.94^{+0.05}_{-0.07}$&$1.27^{+0.09}_{-0.09}$ \\
$\alpha_{\rm X-ray}$\tablenotemark{a}
&$1.61^{+0.36}_{-0.35}$&$4.42^{+1.31}_{-1.01}$&$2.01^{+0.55}_{-0.70}$
&$1.94^{+0.33}_{-0.32}$&$1.76^{+0.35}_{-0.33}$&$2.67^{+0.84}_{-0.68}$&$5.82^{+1.37}_{-1.07}$ \\
$\mathrm{norm}_{\rm powerlaw}$
&$0.64^{+0.24}_{-0.20}$&$0.08^{+0.08}_{-0.05}$&$0.19^{+0.11}_{-0.10}$
&$0.50^{+0.15}_{-0.14}$&$0.45^{+0.15}_{-0.13}$&$0.10^{+0.09}_{-0.07}$&$0.14^{+0.17}_{-0.09}$ \\
$\chi^2/\mathrm{dof}$
&103.74/114&21.79/21&23.3/44&155.44/127&106.98/115&95.63/81&15.9/24 \\
\cutinhead{Derived properties}
$n_{\rm e}$ ($\mathrm{cm^{-3}}$)
&$0.16^{+0.04}_{-0.02}$&$0.14^{+0.02}_{-0.04}$&$0.10^{+0.01}_{-0.02}$
&$0.50^{+0.03}_{-0.03}$&$0.62^{+0.05}_{-0.05}$&$0.53^{+0.05}_{-0.04}$&$0.20^{+0.09}_{-0.05}$ \\
$p_{\rm ther}$ ($\mathrm{keV\,cm^{-3}}$)
&$0.22^{+0.05}_{-0.03}$&$0.19^{+0.04}_{-0.05}$&$0.14^{+0.02}_{-0.03}$
&$0.75^{+0.04}_{-0.03}$&$0.96^{+0.07}_{-0.07}$&$0.81^{+0.05}_{-0.06}$&$0.30^{+0.14}_{-0.11}$ \\
$E_{\rm ther}$ ($10^{53}\ \mathrm{erg}$)
&$8.54^{+1.94}_{-1.16}$&$1.25^{+0.26}_{-0.33}$&$4.33^{+0.62}_{-0.93}$
&$6.78^{+0.36}_{-0.27}$&$2.87^{+0.21}_{-0.19}$&$2.31^{+0.15}_{-0.16}$&$0.96^{+0.42}_{-0.35}$ \\
$L_{\rm X (0.01-100\,keV)}$\tablenotemark{b} ($10^{39}\ \mathrm{erg}$)
&-&-&- &2.78&1.47&1.06&0.25 \\
$L_{\rm H_\alpha +[N II]}$\tablenotemark{b} ($10^{40}\ \mathrm{erg\,s^{-1}}$)
&-&-&-&3.97&2.68&1.21&0.0823 \\
$B$ ($\mathrm{\mu G}$)
&$69.0^{+10.4}_{-10.0}$&$71.8^{+10.6}_{-10.1}$&$66.3^{+10.3}_{-10.0}$
&$109.5^{+14.1}_{-12.9}$&$118.4^{+15.0}_{-13.7}$&$159.3^{+15.9}_{-14.5}$&$90.0^{+11.9}_{-10.9}$ \\
$p_{\rm B}$ ($\mathrm{keV\,cm^{-3}}$)
&$0.12^{+0.04}_{-0.03}$&$0.13^{+0.04}_{-0.03}$&$0.11^{+0.04}_{-0.03}$
&$0.31^{+0.08}_{-0.07}$&$0.35^{+0.09}_{-0.08}$&$0.39^{+0.11}_{-0.09}$&$0.21^{+0.06}_{-0.05}$ \\
$E_{\rm B}$ ($10^{53}\ \mathrm{erg}$)
&$3.11^{+1.04}_{-0.78}$&$0.57^{+0.18}_{-0.13}$&$2.27^{+0.82}_{-0.62}$
&$1.87^{+0.48}_{-0.42}$&$0.70^{+0.19}_{-0.15}$&$0.75^{+0.20}_{-0.16}$&$0.44^{+0.12}_{-0.10}$ \\
$\beta \equiv p_{\rm ther}/p_{\rm B}$
&$1.83^{+1.17}_{-0.65}$&$1.46^{+0.84}_{-0.64}$&$1.27^{+0.73}_{-0.54}$
&$2.42^{+0.87}_{-0.57}$&$2.74^{+1.07}_{-0.72}$&$2.08^{+0.79}_{-0.58}$&$1.43^{+1.32}_{-0.72}$ \\
\enddata
\tablenotetext{a}{For ease of comparison with the radio spectral index $\alpha_{radio}$, we express the X-ray \texttt{powerlaw} slope in the same spectral-index convention, $S_\nu \propto \nu^{-\alpha}$, under the synchrotron interpretation. The fitted photon index $\Gamma$ is therefore converted as $\alpha_{\rm X-ray} = \Gamma - 1$.}
\tablenotetext{b}{The quoted intrinsic luminosities are intended as approximate estimates only. Their values depend sensitively on the adopted absorption/extinction correction, especially for the heavily obscured nuclear regions, and formal uncertainties are therefore not listed.}
\end{deluxetable*}

\subsection{Morphology of the Nuclear Bubbles}\label{sec:morph}

Figure~\ref{fig:north_regions} presents multi-wavelength views of the central region of NGC~4438 in \textit{HST} H$\alpha$, \textit{Chandra} X-rays, VLA $C$-band radio continuum, and radio spectral index. The radio image provides the clearest view of the nuclear outflow morphology. Extended emission is detected on both sides of the nucleus along a northwest--southeast axis, approximately perpendicular to the stellar disk, with a characteristic transverse width of $\sim3\arcsec$ ($\sim200$~pc). On the NW side, the emission forms a bubble-like structure with a projected length of $\sim5\arcsec$ ($\sim350$~pc). This bubble structure is visible in radio, H$\alpha$, and X-rays, and shows an brightened shell. The shell is clear in the radio and H$\alpha$ images, where the brightened regions are spatially coincident. Due to resolution limitations, the substructure of X-rays is not clear. On the SE side, the radio emission extends farther from the nucleus and terminates in a bright shell-like front at a projected distance of $\sim9\arcsec$ ($\sim630$~pc). This front has weaker but spatially corresponding enhancements in H$\alpha$ and X-rays. Interior to this front, the radio image shows a gradually fading, partially shell-brightened structure extending from the nucleus, suggestive of an incomplete or partially obscured bubble. The corresponding H$\alpha$ and X-ray emission from this interior region is much weaker or not clearly detected. If the curvature of the SE shell-like front is extended inward, it connects naturally to the fainter inner radio structure, suggesting that the two SE features are likely parts of the same nuclear bubble whose central portion has low surface brightness or is obscured. The common NW--SE axis also agrees with the orientation of the elongated nuclear radio source and the ionized-gas outflow, consistent with previous studies of the jet direction \citep[e.g.,][]{Hota2007,PuigSubira2026}.

To quantify the visually identified nuclear-bubble structures, we defined subregions based on the multi-wavelength morphology shown in Figure~\ref{fig:north_regions}. The edge-brightened regions are defined as shells, while the fainter regions enclosed by the shells are defined as cavities. The central AGN is excluded using a circular aperture with a radius comparable to the \textit{Chandra} PSF size to minimize contamination from the unresolved nucleus. In the NW bubble, we also define a cap region outside the shell, where the radio emission remains prominent but the H$\alpha$ and X-ray emission are weak. For each band, we measured the mean surface brightness in the shell, cavity, cap, and local background regions, and used the corresponding contrasts and uncertainties to evaluate the significance of the substructures. We also constructed one-dimensional profiles along the common NW--SE axis of the nuclear bubbles using the strip shown in Figure~\ref{fig:north_regions}c. These profiles, shown in Figure~\ref{fig:majoraxis_profile}, provide a complementary view of the surface-brightness variations along the bubble axis. Because the substructures are not strictly one-dimensional, the profile contrasts are diluted, but the main variations remain visible.

The NW nuclear bubble is the most robustly detected structure. Its shell is clearly detected above the local background: the radio surface brightness is higher than the background rms by a factor of $41$, while the H$\alpha$ and X-ray surface brightnesses are higher than the corresponding local background levels by factors of $2$ and $273$, respectively. The radio emission is also strongly edge-brightened. The shell is 75\% brighter than the cavity, and the shell--cavity difference corresponds to $17\sigma$. The H$\alpha$ image shows a similar shell--cavity morphology, with the enhanced emission spatially coincident with the radio shell. The X-ray emission is broadly associated with the NW bubble, although the internal shell--cavity substructure is not clearly resolved, likely because of limited photon statistics and effective resolution. The NW cap differs from the shell mainly through its band dependence: its radio surface brightness is about 43\% of the shell value, whereas its X-ray and H$\alpha$ surface brightnesses are only about 8\% and 17\% of the cavity values, respectively. The radio spectral-index map also separates the shell from the other NW subregions: the shell has a flatter mean spectral index, $\alpha_{\rm radio}=0.94^{+0.05}_{-0.07}$, while the cavity and cap have steeper values of $\alpha_{\rm radio}=1.23\pm0.07$ and $1.27\pm0.09$, respectively. We therefore define the cap as a radio-bright structure outside the multi-wavelength shell, and use ``cavity'' only as a morphological label for the fainter interior enclosed by the shell.

The SE nuclear bubble is less complete than the NW bubble, but the outer shell-like front is still supported by the regional photometry. In the radio band, the SE shell has a mean surface brightness of $\sim4\,\sigma_{\rm rms}$ above the background. In X-rays, the same shell region is more than 16 times brighter than the local background level. The H$\alpha$ image shows localized enhancements along the outer front, although the emission is weaker and less continuous than on the NW side. We therefore identify this structure as the SE nuclear bubble, while treating its shell--cavity decomposition as less secure than that of the NW bubble.

A useful comparison can be made with the nuclear superbubble in NGC~3079, the only external galaxy in which diffuse hard X-ray synchrotron emission has been reported on bubble scales \citep{Li2019}. 
Although the superbubbles in NGC~3079 are much larger (diameters of $\sim1.1$--$1.5$ kpc) than the $\sim200$ pc nuclear bubbles in NGC~4438, both systems show a similar morphological combination of limb-brightened shell/cavity structures and an additional radio-bright feature located ahead of or above the main shell.
In both galaxies, this offset radio-bright component is more prominent in the radio continuum than in H$\alpha$ or soft X-rays. 
This resemblance suggests that the northern cap in NGC~4438 is not a unique morphological feature, although its physical origin is discussed further in Section~\ref{sec:cap}.

\subsection{Radio and X-ray Spectral Modeling of the Nuclear Bubbles}\label{sec:spectral}

Diffuse hard X-ray emission is detected from the northwest nuclear bubble and exhibits an extended morphology on scales comparable to the bubble itself (Section~\ref{sec:morph}), ruling out the possibility of contamination from unresolved point sources.
Previous studies have explored various spectral models for similar systems and have shown that a non-thermal synchrotron component provides the most adequate description of the hard X-ray emission \citep[e.g.,][]{Li2022, Li2019}.
In particular, \cite{Li2022} demonstrated that a thermal-plus-power-law model outperforms purely thermal models, such as two-temperature \texttt{apec} models, in reproducing the hard X-ray emission. 
At radio frequencies, the steep spectrum of the NW bubble and the flatter nuclear component are broadly consistent with the high-resolution radio analysis of \citet{PuigSubira2026}, supporting a synchrotron origin for the extended bubble emission.
Following \citet{Li2022} and \citet{PuigSubira2026}, we adopt a framework in which the X-ray spectrum consists of thermal plasma emission plus a hard synchrotron power-law component, while the radio emission is produced by synchrotron radiation. We extend these studies by testing whether the radio and X-ray synchrotron emission can arise from a single electron population and by performing spatially resolved fitting of the shell, cavity, and cap regions revealed by the new high-resolution VLA data (Figure~\ref{fig:north_regions}a). Although this subdivision reduces the X-ray photon statistics in some regions, the model is determined from the integrated bubble spectra and previous studies and is applied uniformly to the subregions to examine spatial differences rather than selected independently for each low-count spectrum.

\begin{figure*}[!t]
\centering
\includegraphics[width=\textwidth]{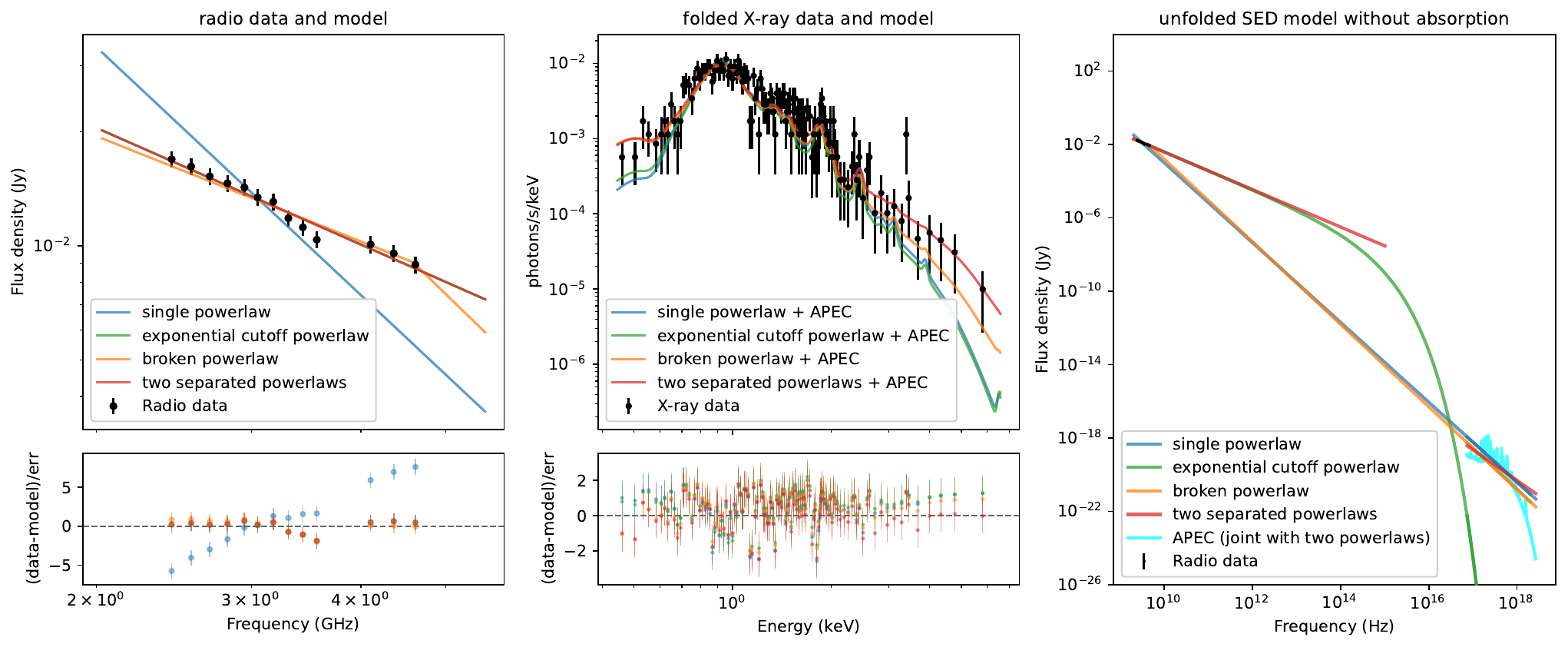}
\caption{
Illustration of different models for the joint radio and X-ray spectral fitting, shown here for the cavity region as an example.
The left and middle panels show the radio and X-ray spectra, respectively.
Black points denote the data, while the blue, green, orange, and red curves show the single power-law, exponential cutoff power-law, broken power-law, and two separate power-law models.
The lower panels show the residuals.
The right panel shows the broadband spectral energy distribution (SED), in which the synchrotron and thermal components are displayed separately; only one representative thermal model is shown, in cyan.
A key feature of the data is that both the radio spectral index ($\alpha_{\rm radio}\sim1$) and the X-ray spectral index ($\alpha_{\rm X\mbox{-}ray}\sim2$) are relatively flat, whereas the effective slope connecting the radio and X-ray bands is much steeper ($-[ln(f_{X-ray})-ln(f_{radio})]/[ln(\nu_{X-ray})-ln(\nu_{radio})]>3$)). As a result, single power-law could not fit the data, the broken power-law and cutoff power-law models both leave residuals in the X-ray band, indicating that two separate power-law components are required to describe the radio and hard X-ray emission.
}
\label{fig:cavity_joint_radio_xray_comparison}
\end{figure*}

In practice, we first attempted a joint fit in which the radio and X-ray synchrotron emission within the same spatial region were described by a single electron population.
In this approach, the X-ray spectrum was modeled as the sum of thermal emission and a synchrotron power-law component, while the radio spectrum was assumed to arise solely from the same non-thermal component.
Starting from a simple power-law description, we further explored models with an exponential cutoff and a broken power-law electron spectrum.
However, none of these models can simultaneously reproduce the radio spectrum and improve the X-ray fit relative to a single-\texttt{apec} model; all of them leave obvious residuals in the hard X-ray band, shown in Figure~\ref{fig:cavity_joint_radio_xray_comparison}.
The key difficulty is that, if the radio synchrotron component is extrapolated to the X-ray band as a single power law, its predicted normalization exceeds the observed hard X-ray component by approximately 8 orders of magnitude, even though their spectral slopes differ only modestly.
As a result, models that adequately fit the radio data inevitably overpredict the hard X-ray emission, whereas models tuned to the X-ray band contribute negligibly at radio frequencies.
We are therefore compelled to relax the assumption of a single electron population and instead model the radio and X-ray synchrotron components separately, while retaining a power-law description for each.

\begin{figure*}
\centering
\includegraphics[width=0.48\textwidth]{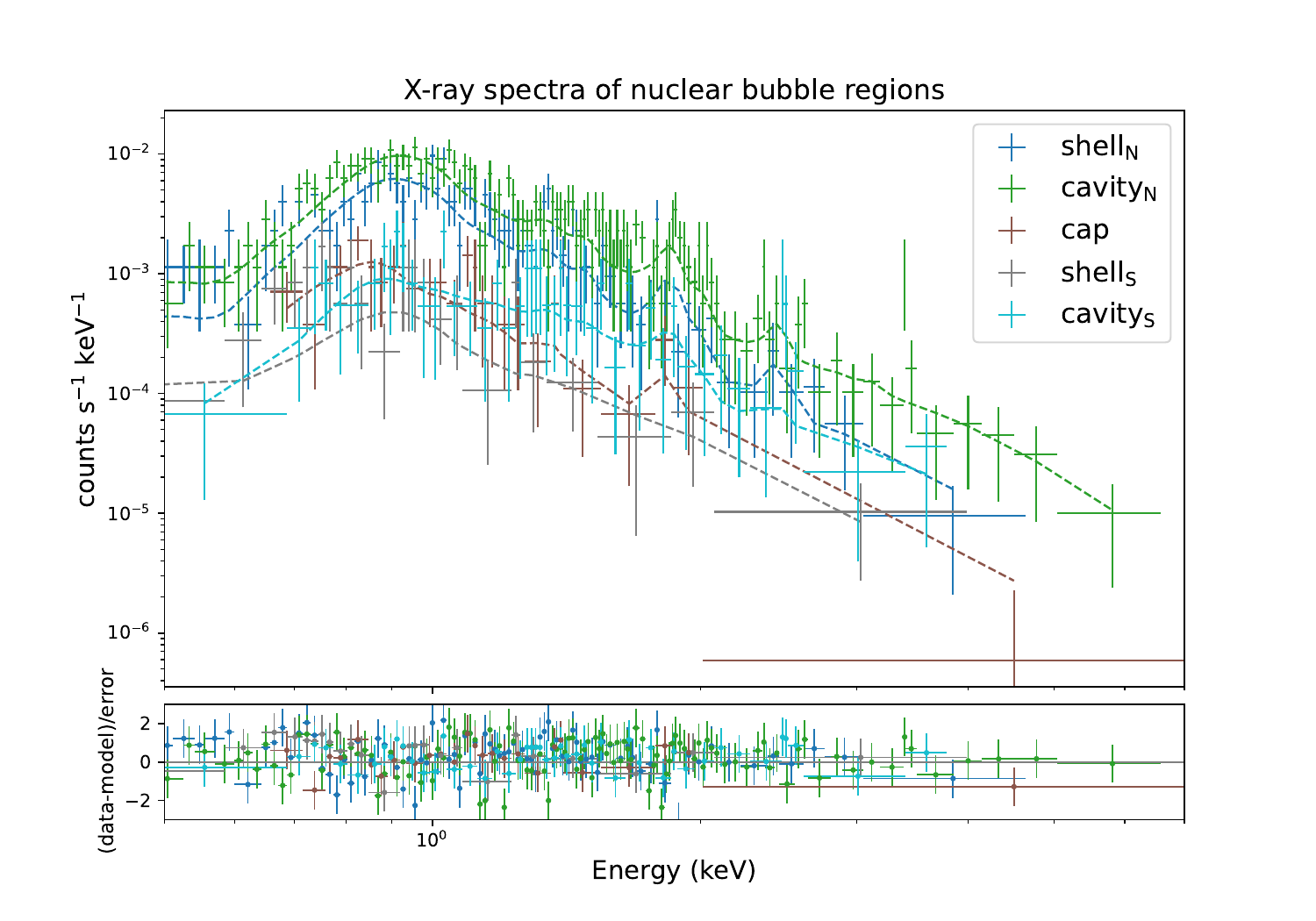}
\includegraphics[width=0.48\textwidth]{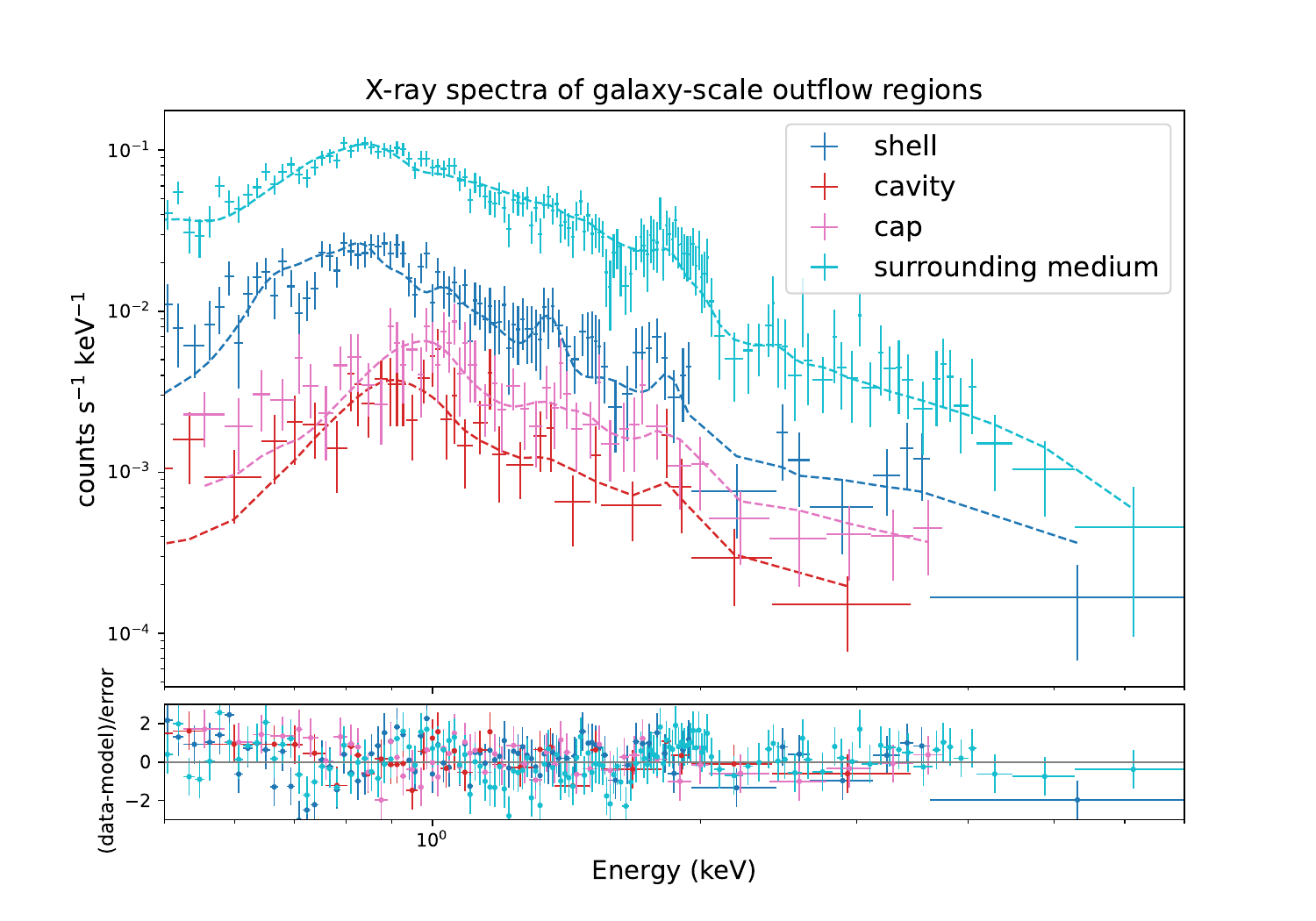}
\caption{
\textit{Chandra} X-ray spectra and best-fit models for the main extraction regions.
Panel (a) shows the spectra of the nuclear-bubble regions, including the NW shell, NW cavity, cap, SE shell, and SE cavity.
Panel (b) shows the spectra of the galaxy-scale outflow regions, including the shell, cavity, cap, and surrounding medium.
In each panel, the data points show the observed spectra, the dashed curves show the best-fit TBabs$\times$mxabs$\times$(apec+power-law) models, and the lower subpanel shows the residuals in units of $(\mathrm{data}-\mathrm{model})/\mathrm{error}$.
The corresponding best-fit parameters are listed in Tables~\ref{tab:fitting_small_bubble} and \ref{tab:fitting_large_bubble}.
}
\label{fig:xray_spectra_all}
\end{figure*}

Therefore, we model the X-ray spectra using the form \texttt{TBabs} $\times$ \texttt{mxabs} $\times$ (\texttt{apec} + power-law), where the \texttt{apec} component describes the thermal plasma dominating the soft X-ray band, and the power-law component represents synchrotron emission.
The Galactic absorption is fixed at $N_{\rm H,MW}=2.13\times10^{20}\,\mathrm{cm^{-2}}$, while an additional intrinsic absorption component is included.
Metal abundances and redshift are fixed in all regions.
The best-fit parameters for different spatial regions are summarized in Table~\ref{tab:fitting_small_bubble}. 
Figure~\ref{fig:xray_spectra_all} shows the X-ray spectra, best-fit models, and residuals for the main extraction regions used in both the nuclear-bubble and galaxy-scale analyses.

Due to limited photon statistics in individual subregions, several parameters are fixed or linked to reduce degeneracies.
Within each bubble, the gas temperature for the shell and cavity is fixed to the value obtained for the corresponding whole-bubble fit.
Similarly, the intrinsic absorption for shell and cavity regions within the same bubble is linked.
These choices allow robust comparisons of spectral properties between regions while maintaining statistically meaningful constraints.

The thermal component shows only modest variation across different regions.
In both bubbles, the best-fit temperatures are clustered around $kT \sim 0.7$--$0.8$~keV, with no statistically significant temperature difference between shell and cavity.
The southern bubble is marginally cooler than the northern one, but the overall thermal properties are remarkably uniform.

In contrast, the non-thermal radio spectral indices exhibit clear spatial trends. 
Broadly consistent with the high-resolution radio results of \citet{PuigSubira2026}, our finer shell/cavity/cap decomposition shows that in both bubbles the radio spectrum is flattest at the shell. 
In the northern bubble, the shell has $\alpha_{\rm radio}=0.94^{+0.05}_{-0.07}$, significantly flatter than both the cavity ($\alpha_{\rm radio}=1.23^{+0.07}_{-0.07}$) and the radio cap ($\alpha_{\rm radio}=1.27^{+0.09}_{-0.09}$). 
A similar trend is present in the southern bubble, where the shell ($\alpha_{\rm radio}=1.11^{+0.07}_{-0.06}$) is flatter than the cavity ($\alpha_{\rm radio}=1.23^{+0.07}_{-0.07}$).

The X-ray non-thermal component shows the opposite spatial behavior.
In the northern bubble, the cavity exhibits the flattest X-ray spectrum, $\alpha_{\rm X-ray}=1.76^{+0.35}_{-0.33}$, while the shell is steeper ($\alpha_{\rm X-ray}=2.67^{+0.84}_{-0.68}$), and the radio cap is steepest ($\alpha_{\rm X-ray}=5.82^{+1.37}_{-1.07}$).
In addition, the power-law normalization per unit area in the cavity is several times higher than in the shell and cap, indicating that the hard X-ray synchrotron emission is centrally concentrated.
In the southern bubble, the X-ray spectral indices follow a broadly similar trend, although the uncertainties are larger owing to the weaker X-ray emission.

In summary, while the thermal properties show little systematic variation between regions, the non-thermal radio and X-ray components display pronounced and opposite spatial trends.
The radio spectrum is flattest at the shell, whereas the X-ray spectrum is flattest and strongest in the cavity. 

\subsection{The Galaxy-scale Outflow}\label{sec:outflow}

\begin{deluxetable*}{lccccc}
\tablecaption{Spectral fitting parameters and derived properties of the galaxy-scale outflow.\label{tab:fitting_large_bubble}}
\tablewidth{0pt}
\tablehead{
\colhead{Parameter} & \colhead{disk} & \colhead{cavity} & \colhead{shell} & \colhead{cap} & \colhead{surrounding medium}
}
\startdata
area ($arcmin^2$) &1.01 &0.45 &3.78 &0.82 &28.39 \\
$N_{\rm H,MW}$ ($10^{22}\,\mathrm{cm^{-2}}$)      
    & \multicolumn{5}{c}{0.0213 (fixed)} \\
$N_{\rm H,int}$ ($10^{22}\,\mathrm{cm^{-2}}$)     
 & $0.05^{+0.20}_{-0.04}$& -& -&-& -\\
kT (keV)     
    & $0.63^{+0.04}_{-0.10}$& $0.81^{+0.07}_{-0.07}$&$0.64^{+0.03}_{-0.03}$&$0.97^{+0.08}_{-0.08}$  & $0.62^{+0.02}_{-0.02}$\\
$Z$ ($Z_\odot$)   
    & \multicolumn{4}{c}{1.0 (fixed)}& $0.05^{+0.01}_{-0.01}$\\
$\mathrm{norm}_{\rm apec}$ ($10^{-5}$)
   &$1.55^{+0.77}_{-0.17}$&$0.41^{+0.06}_{-0.06}$& $3.07^{+0.14}_{-0.14}$&$0.71^{+0.13}_{-0.11}$& $162.4^{+10.4}_{-10.3}$\\
$\Gamma$
  & \multicolumn{5}{c}{1.4(fixed)} \\
$\mathrm{norm}_{\rm powerlaw}$ 
    & $0.46^{+0.09}_{-0.09}$& $0.22^{+0.05}_{-0.05}$&  $1.20^{+0.12}_{-0.13}$&$0.54^{+0.08}_{-0.08}$& $4.03^{+0.46}_{-0.46}$\\
$\chi^2/\mathrm{dof}$
  & 60.55/59& 25.31/31 & 110.20/84 & 39.01/49 & 109.68/135\\
\cutinhead{Derived properties}
$n_{\rm e}$ ($\mathrm{cm^{-3}}$)& $0.0032^{+0.0008}_{-0.0001}$&$0.0025^{+0.0002}_{-0.0002}$&$0.0024^{+0.0001}_{-0.0001}$&$0.0024^{+0.0003}_{-0.0002}$& $0.0036^{+0.0001}_{-0.0001}$\\
$p_{\rm ther}$ ($\mathrm{keV\,cm^{-3}}$)&$0.0040^{+0.0005}_{-0.0003}$& $0.0039^{+0.0005}_{-0.0005}$& $0.0029^{+0.0001}_{-0.0001}$&$0.0045^{+0.0008}_{-0.0007}$&$0.0043^{+0.0001}_{-0.0001}$\\
 $E_{\rm ther}$ ($10^{56}\ \mathrm{erg}$)&$0.42^{+0.05}_{-0.03}$ & $0.18^{+0.02}_{-0.02}$& $1.15^{+0.05}_{-0.05}$& $0.39^{+0.06}_{-0.06}$&$38.3^{+1.3}_{-1.3}$\\
  $L_{\rm X (0.01-100.0\ keV)}$ ($10^{40}\ \mathrm{erg/s}$)&- & $1.94$& $6.71$& $4.42$&-\\
$B$ ($\mathrm{\mu G}$)
   & -&$12.43^{+1.70}_{-1.67}$&$10.4^{+1.46}_{-1.35}$ &$9.65^{+1.39}_{-1.27}$& -\\
$p_{\rm B}$ ($\mathrm{keV\,cm^{-3}}$)
   &-& $0.004^{+0.001}_{-0.001}$& $0.0028^{+0.0008}_{-0.0007}$ &$0.0023^{+0.0007}_{-0.0006}$& -\\
$\beta \equiv p_{\rm ther}/p_{\rm B}$
   & - & $0.98^{+0.49}_{-0.30}$ & $1.04^{+0.39}_{-0.26}$ & $1.96^{+1.16}_{-0.69}$ & -
\enddata
\end{deluxetable*}

Figure~\ref{fig:outflow} presents a multi-wavelength view of the galaxy-scale outflow in NGC~4438. Previous optical and H$\alpha$ studies (e.g., \citealt{Kenney2008}) have shown that NGC~4438 is undergoing strong tidal interaction within the Virgo cluster environment. 
The H$\alpha$ image reveals numerous filamentary structures extending toward M86, widely interpreted as gas stripped from the disk through galaxy--galaxy interaction.

However, the large-scale radio morphology reveals a prominent, coherent bubble-like structure extending above the disk.
Such a symmetric, collimated radio bubble is unlikely to arise from tidal stripping alone.
Given the relatively low star formation rate of NGC~4438 ($\lesssim 0.5\,M_\odot\,{\rm yr^{-1}}$; e.g., \citealt{Boselli2016}), stellar feedback is also insufficient to account for the observed radio luminosity and spatial extent.
We therefore interpret the radio bubble as the product of an AGN-driven outflow rather than interaction-induced star formation or tidal processes.

The extended H$\alpha$ filaments outside the bubble, shown by the yellow polygon in Figure~\ref{fig:outflow}, are likely dominated by interaction-driven stripped gas; we define these outer regions as the \emph{filament} component and exclude them from the AGN energetic analysis.

Within the bubble, we divide the structure into three physically motivated subregions.
First, the outer rim where H$\alpha$, radio, and X-ray emission are all prominent is defined as the \emph{shell}.
The spatial correspondence between H$\alpha$ and X-ray emission in similar systems has been discussed previously (e.g., \citealt{Machacek2004}), and here the multi-phase emission suggests shock compression and heating along the bubble boundary.
Second, the interior of the bubble contains two regions with reduced H$\alpha$, X-ray, and radio surface brightness, which we define as the \emph{cavity}.
Finally, toward the outermost edge of the bubble, we identify a region where the radio emission remains strong while the H$\alpha$ and X-ray emission weaken significantly; following our terminology for the northwest nuclear bubble, we refer to this feature as the \emph{cap}.
In addition, the blue region in Figure~\ref{fig:outflow} corresponds to H$\alpha$ emission associated with the galactic disk and is defined as the \emph{disk} component.
In the following contents, we concentrate on the cavity, shell, and cap regions of the radio bubble, where the AGN-driven outflow is expected to dominate the energetics.

Spectra are extracted from each region, and joint spectral modeling is performed using the model \texttt{TBabs} $\times$ \texttt{mxabs} $\times$ (\texttt{apec} + \texttt{powerlaw}).
The best-fit spectral parameters are summarized in Table~\ref{tab:fitting_large_bubble}.
The power-law component is included to account for unresolved point-like sources within the extraction regions. The spectra and best-fit models for these galaxy-scale regions are included in Figure~\ref{fig:xray_spectra_all}, together with the nuclear-bubble spectra for comparison.

In the galaxy-scale outflow, neither the thermal gas properties nor the magnetic field strength show strong systematic variations among the cavity, shell, and cap regions.
Within uncertainties, the temperature and magnetic field strength remain broadly comparable across these substructures, indicating that the 10-kpc scale outflow is relatively homogeneous on these scales.

In contrast to the nuclear bubbles, however, the physical conditions in the galaxy-scale outflow are markedly different. 
The characteristic temperature decreases from $kT\sim0.8$~keV in the nuclear bubbles to $kT\sim0.12$~keV in the galaxy-scale outflow, and the magnetic field strength drops from $\gtrsim100~\mu{\rm G}$ to the $\sim10~\mu{\rm G}$ level.
Despite this overall decline in energy density from sub-kiloparsec to kiloparsec scales, the relative contribution of magnetic pressure becomes slightly larger. 
Nevertheless, the thermal and magnetic pressures remain comparable within uncertainties, implying that the galaxy-scale outflow is jointly supported by thermal gas and magnetic fields.

\section{Discussion}\label{sec:discussion}

\subsection{Origin of the Non-thermal Emission in the Nuclear Bubbles}

Diffuse hard X-ray synchrotron emission on bubble scales has been reported in only one other external galaxy, NGC~3079 \citep{Li2019}. 
This makes NGC~3079 the most relevant comparison system for interpreting the non-thermal X-ray emission in NGC~4438. 
However, as we show below, the compact nuclear bubbles in NGC~4438 in NGC~4438 not only provide another candidate case of bubble-scale high-energy particle acceleration, but also reveal a more complex spatial relationship between the radio- and X-ray-emitting electrons.

We interpret the power-law component detected in the hard X-ray band as synchrotron radiation from extremely high-energy electrons ($\sim {\rm TeV}$), following \citet{Li2022}, who showed that a thermal plus power-law model provides a significantly better description of the hard X-ray emission than purely thermal models such as two-temperature \texttt{apec} representations. 
This interpretation is further supported by the diffuse morphology of the X-ray emission on the scale of the northwest nuclear bubble, which shows little contamination from unresolved point sources (Section~\ref{sec:morph}). 
Although alternative emission mechanisms cannot be entirely ruled out, synchrotron radiation from very energetic electrons offers the most natural explanation for the observed spectral and spatial properties.

A key physical implication of this interpretation is the extremely short cooling timescale of the X-ray-emitting synchrotron electrons. 
Under a magnetic field of order $100~\mu{\rm G}$, the GeV electrons responsible for radio synchrotron emission cool on Myr timescales, whereas electrons emitting hard X-ray synchrotron radiation must reach multi-TeV energies and cool on $\lesssim 10^2$~years \citep[see][]{Li2022}, much shorter than the dynamical timescale of the bubble (Section~\ref{sec:energy_budget_baby}). 
If the hard X-ray power-law component is indeed synchrotron radiation, continuous in-situ acceleration is therefore required. 
The X-ray-emitting electrons in the cavity cannot simply represent a long-lived fossil population accelerated at a distinct shock or passively transported from the nucleus. 

Two additional observational findings further suggest that the electrons responsible for the radio and X-ray emission cannot be characterized as a single, continuous electron population shaped by one uniform acceleration history. 
First, the combined radio and X-ray spectral analysis described in Section~\ref{sec:spectral} demonstrates that the non-thermal emission linked to the northwest nuclear bubble in NGC~4438 cannot be accounted for by a single, broadband electron energy distribution. 
In particular, if the radio-emitting electrons are extrapolated as a single power-law population, the normalization required to match the hard X-ray emission differs by approximately eight orders of magnitude in the synchrotron interpretation. 
Even if the hard X-rays are instead attributed to inverse-Compton scattering, the predicted flux would still exceed the observed level by about five orders of magnitude when considering only the cosmic microwave background as the target photon field. 
These discrepancies, together with the failure of single power-law and exponentially cutoff synchrotron models, argue strongly against a single electron population extending smoothly from the radio to the hard X-ray band.

Second, the spatial trends of the radio and X-ray spectral indices are reversed. 
In the radio band, the spectral index steadily steepens when moving from the shell toward the galactic center, with the flattest spectrum observed at the shell (see the bottom-right panel of Figure~\ref{fig:north_regions} and Figure~\ref{fig:majoraxis_profile}). 
This region spatially coincides with the shock front identified from the shell--cavity morphology. 
In contrast, the X-ray power-law component is spectrally harder in the cavity and becomes steeper toward the shell and cap. 
Moreover, the power-law normalization per unit area in the cavity is three to four times larger than in the shell and cap regions (Table~\ref{tab:fitting_small_bubble}), indicating that the non-thermal hard X-ray emission is dominated by the cavity region rather than by the shell.

Together, these results suggest a two-component picture. 
The GeV electron population responsible for the radio synchrotron emission is most naturally associated with shock acceleration at the expanding shell, followed by radiative aging as the electrons advect inward. 
In this picture, the shock is launched near the AGN and propagates outward through the surrounding medium. 
Electrons accelerated at earlier epochs, which now reside closer to the AGN inside the cavity, have experienced longer synchrotron and inverse-Compton cooling, leading to a steeper radio spectrum. 
By contrast, electrons at the present-day shock front, corresponding to the shell region, are freshly accelerated and therefore exhibit the flattest radio spectral index. 
This scenario naturally explains both the spatial ordering of the radio spectral index and the limb-brightened shell morphology observed in the radio continuum.

However, the hard X-ray synchrotron component requires an additional high-energy acceleration channel operating locally within the bubble interior. 
Since the relevant electrons cool rapidly, any viable mechanism must continuously inject or re-accelerate a small fraction of particles to TeV energies in situ, distinct from the shell-related radio acceleration. 
A particularly plausible possibility is magnetic reconnection associated with the AGN outflow. 
In magnetically dominated inner jets, reconnection can efficiently convert magnetic energy into non-thermal particles and produce hard electron spectra (e.g., \citealt{SironiSpitkovsky2014,Guo2014,Giannios2013}). 
In the present system, such acceleration may plausibly occur in localized regions linked to the inner outflow and cavity environment, including jet--cloud interaction sites, shear layers, or other magnetically stressed structures within the bubble. 
Additional possibilities include direct electric-field acceleration in magnetospheric gap-like regions near the black hole poles (e.g., \citealt{Levinson2000,RiegerAharonian2008}), intermittent internal shocks driven by variability in the nascent outflow (e.g., \citealt{Spada2001,BlandfordKonigl1979}), and second-order Fermi acceleration powered by turbulence within the cavity. 
The latter is qualitatively supported by the enhanced velocity dispersion measured in the cavity in H$\alpha$ (see Figure~9 in \citealt{Hermosa2024}). 
Although the current data do not allow us to distinguish uniquely among these scenarios, they strongly suggest that the hard X-ray synchrotron component requires ongoing local particle acceleration distinct from the process that produces the shell-dominated radio synchrotron emission.

\subsection{The Nature of the Radio Cap}\label{sec:cap}

\begin{figure}
    \centering
    \includegraphics[width=1.0\linewidth]{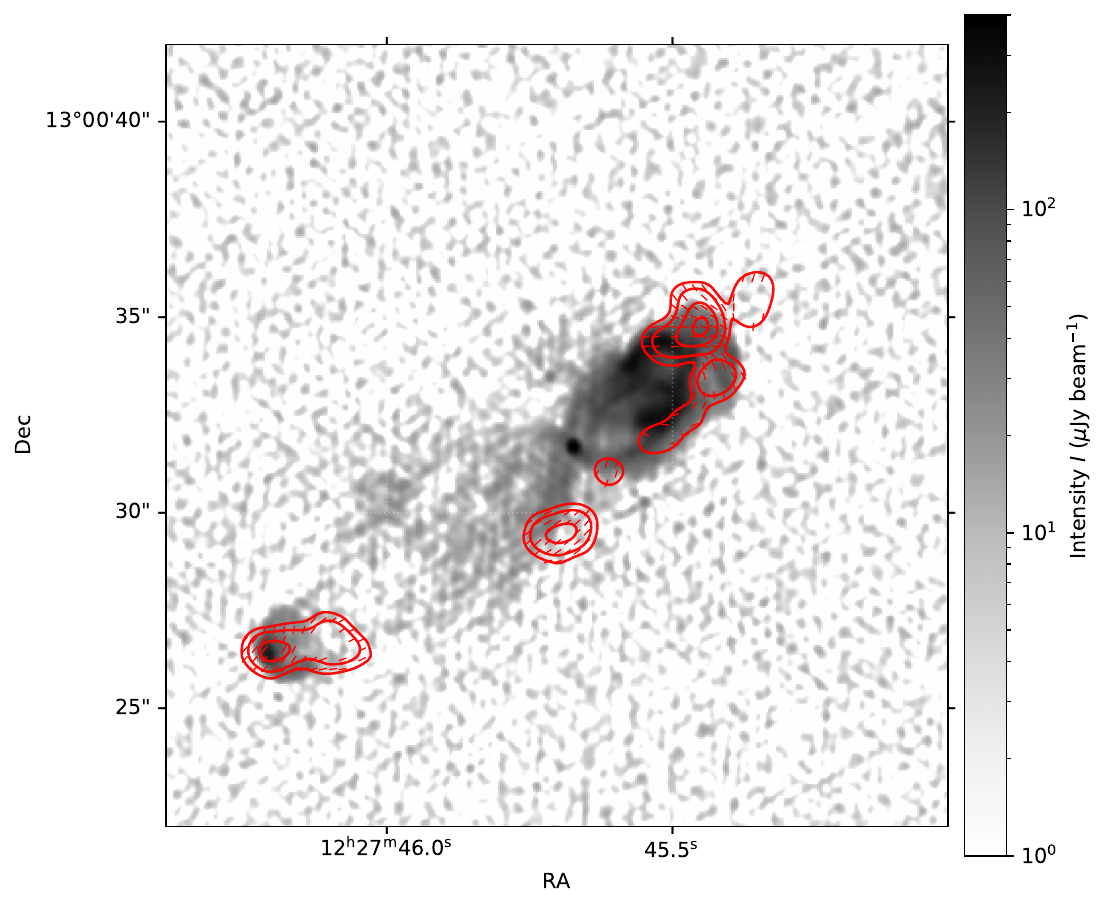}
    \caption{
    Radio continuum total-intensity image of the central region of NGC~4438,
    shown in grayscale, with contours of linearly polarized intensity overlaid in red.
    Polarization vectors, indicating the inferred magnetic-field orientations, are also shown.
    The rms noise of the polarization map is $\sigma_P = 6.0\,\mu{\rm Jy\,beam}^{-1}$,
    and contours are drawn at 1.5, 2.0, 3.0, and 4.0 $\sigma_P$. Contour segments for which the mean Stokes~$I$ intensity along the segment is below $2\sigma_I$ are omitted.
    The polarized emission is detected primarily in the radio-bright cap region,
    while no significant polarized emission is detected above the $6\sigma_P$ level.
    }
\label{fig:pol_contour}
\end{figure}

A distinctive feature of the northwest nuclear bubble in NGC~4438 is the presence of a radio-bright cap located ahead of the shell. 
This region is characterized by prominent radio synchrotron emission, while the corresponding H$\alpha$ and X-ray emission are weak or absent (Section~\ref{sec:morph}). 
This spatial separation between radio and thermal tracers indicates that the radio cap represents a physically distinct component of the system rather than a simple extension of the shell or cavity emission and provides the clearest example of this phenomenon among the compact bubbles in NGC~4438.

An additional observational signature of the northern radio cap is the detection of linearly polarized emission (Figure~\ref{fig:pol_contour}), which is not observed in the northern shell or cavity regions.
The presence of polarized synchrotron emission implies a relatively ordered magnetic field configuration within the cap.
Although the polarization fraction is modest, it nevertheless indicates that the magnetic field in the cap region remains dynamically significant.
Consistent with this interpretation, the magnetic field strength inferred for the radio cap is comparable to that measured in the shell, despite the lack of strong
thermal emission at the same location.

The comparison with NGC~3079 is particularly instructive in this context. 
In that system, the radio emission is likewise weak inside the H$\alpha$/soft-X-ray bubble but becomes enhanced in a cap-like structure above the bubble, indicating a clear spatial offset between the radio and thermal tracers \citep{Li2019}. 
Importantly, the radio cap in NGC~3079 is also strongly polarized, suggesting that it traces a region of enhanced and relatively ordered magnetic field \citep{Li2019}. 
Our northern cap in NGC~4438 appears to follow the same qualitative pattern but on a much smaller scale: it is brightest in the radio, lacks comparably strong H$\alpha$ and X-ray counterparts, and is the only location in the northern bubble where linearly polarized emission is clearly detected (Figure~\ref{fig:pol_contour}). 
Although the polarization signal in NGC~4438 is more modest than that reported for NGC~3079, the resemblance between the two systems strengthens the interpretation that radio caps are magnetically distinct structures associated with the leading edge of expanding jet-driven bubbles. 
This picture is also consistent with the energetic analysis of the NGC~3079 superbubble, in which the radio cap extends beyond the thermal bubble and the magnetic field is argued to remain dynamically important even where the thermal emission becomes faint \citep{Li2024}. 
Taken together, these similarities suggest that radio caps may represent a generic MHD response to bubble expansion rather than a peculiarity of any individual system. 

A plausible physical interpretation of such a radio cap is magnetic draping, a process in which ambient magnetic field lines are swept up and wrapped around the leading surface of a moving or expanding obstacle in a magnetized plasma \citep{Lyutikov2006,DursiPfrommer2008}. 
In the context of jet-driven bubbles, draping can form a thin layer of an enhanced, relatively ordered magnetic field along the bubble boundary, helping suppress mixing and making the interface more sharply defined. 
This provides a natural framework for understanding why the northern cap in NGC~4438 is bright in synchrotron emission, in front of a thermally bright shell, and is the only location in the northern bubble where linearly polarized emission is clearly detected. 
Similar draping-like field geometries have been invoked to explain banded Faraday rotation measure (RM) structures, with iso-RM contours oriented roughly perpendicular to the radio lobe axes, and other anisotropic magnetoionic structures around radio lobes in several radio galaxies \citep[e.g.,][]{Guidetti2011,Guidetti2012} and have also been suggested as an explanation for polarized structures in AGN lobes such as B2~0258 + 35 \citep{Adebahr2019}. 
The qualitative resemblance between these systems and the northern cap in NGC~4438 suggests that magnetic draping may contribute to shaping the radio-bright leading edge of the bubble. 

By contrast, the southern bubble does not show a comparably clear radio-bright feature offset from the thermally bright shell. 
Instead, the radio, H$\alpha$, and X-ray emission on the southern side remain more broadly spatially coherent, resembling the more common multi-wavelength shell/tip morphology rather than the distinct offset cap seen in the north. 
This asymmetry may reflect differences in the ambient environment on the two sides of the nucleus. 
If the northern bubble expands into denser gas, as also suggested by the dynamical arguments presented in Section~\ref{sec:energy_budget_baby}, a draping-like magnetized layer at its leading edge may be easier to develop or preserve. 
The southern bubble, by contrast, may remain dominated by a more spatially coherent shell structure.

On larger scales, the 10-kpc-scale outflow in NGC~4438 also contains regions of
enhanced radio emission with weak or undetected H$\alpha$ and X-ray counterparts (`cap' in Table~\ref{tab:fitting_large_bubble}).
The `cap' in the galaxy-scale outflow may reflect a breakout or partial rupture of the bubble boundary rather than a classical magnetized leading-edge cap.
In this picture, the thermal shell becomes disrupted or strongly diluted in this direction, causing the associated H$\alpha$ and X-ray emission to fade, while the synchrotron-emitting relativistic particles and magnetic field remain visible as a smoother and more extended radio structure. 
Such a scenario would naturally explain the absence of clear thermal counterparts in the cap region, as well as its relatively smooth radio morphology.

\subsection{Energetics of Jet-driven Bubbles across Spatial Scales}

A key open question in AGN feedback is how the mechanical energy carried by jets is redistributed among kinetic, thermal, magnetic, and radiative components as the outflow propagates away from the nucleus. 
NGC~4438 provides a particularly useful laboratory for addressing this question, because it hosts both compact nuclear bubbles and a much larger $\sim10$ kpc galaxy-scale outflow that is plausibly powered by the same central engine.

The energetics of the nuclear bubbles have already been explored in previous work. Using \textit{Chandra} data, \citet{Li2022} characterized the thermal properties of the hot X-ray–emitting gas, while high-resolution radio observations by \citet{PuigSubira2026} provided important constraints on the non-thermal components and discussed the jet power using empirical radio-based scalings. 
In this work, we build on these studies in two more specific ways. 
First, we derive a spatially resolved energy budget for individual bubble substructures, including the shell, cavity, and cap regions, rather than considering only the bubbles as a whole. 
Second, instead of inferring the overall energetics primarily from single-waveband proxies or empirical jet-power relations, we estimate the thermal, kinetic, magnetic/cosmic-ray, and radiative components directly from the X-ray, H$\alpha$, and radio data and then combine them into a unified energy census.

Our main goal in this subsection is therefore twofold. 
We first derive a resolved multi-wavelength energy budget for the nuclear bubbles and compare it with the more commonly adopted $4pV$-based and empirical radio--jet power estimates. 
We then extend the same energetic reasoning to the galaxy-scale outflow, in order to explore how the dominant energy reservoirs vary with spatial scale and whether the nuclear bubbles and the galaxy-scale outflow can be understood within a common feedback framework.

\subsubsection{Resolved Energy Budget of the Nuclear Bubbles}
\label{sec:energy_budget_baby}

We now use the multi-wavelength data to estimate the magnitude and partition of the different forms of energy in the nuclear bubbles, including the thermal, kinetic, magnetic/cosmic-ray, and radiative components. 
Table~\ref{tab:fitting_small_bubble} summarizes the relevant physical quantities derived from the multi-wavelength data. 
In particular, X-ray spectral fitting leads to the estimation of the electron density, thermal pressure, thermal energy, and X-ray luminosity of the hot gas; the H$\alpha$ image constrains the luminosity of the warm ionized component; and the radio data provide estimates of the magnetic field strength, magnetic pressure, and magnetic energy. 
The projected areas of the cavity, shell, and cap regions are measured directly from the imaging data and, together with simple geometric assumptions, are used to estimate the corresponding volumes.

We begin with the northwest nuclear bubble, which is morphologically cleaner and better constrained at all wavelengths. 
To estimate its dynamical timescale, we use the density jump between the cap and shell to infer the shock Mach number, treating the cap as the pre-shock region and the shell as the post-shock region:
\begin{equation}
r \equiv \frac{n_2}{n_1}
= \frac{n_e(\mathrm{shell_N})}{n_e(\mathrm{cap})}
= \frac{0.53}{0.20} = 2.65.
\end{equation}
For $\gamma=5/3$, the Rankine--Hugoniot relation gives $\mathcal{M}=2.43$. 
Using $kT_{\rm shell}\simeq0.80$~keV (Table~\ref{tab:fitting_small_bubble}) and $\mu=0.61$, we obtain a sound speed of $c_s\simeq458$~km~s$^{-1}$, implying a shock velocity of $v_{\rm sh,N}\simeq1.11\times10^3$~km~s$^{-1}$.

Adopting the projected distance from the AGN to the northern shock front as $R_{\rm proj,N}=4\arcsec$, and correcting for the galaxy inclination ($i\sim80^\circ$; \citealt{Kenney2008}) assuming that the jet is approximately perpendicular to the disk, we estimate an intrinsic distance
\begin{equation}
R_N=\frac{R_{\rm proj,N}}{\sin i},
\end{equation}
which gives $R_N\simeq4.1\arcsec$ ($\simeq285$ pc). 
The corresponding dynamical time,
\begin{equation}
t_{\rm dyn}=\frac{R_N}{v_{\rm sh}},
\end{equation}
is then $t_{\rm dyn}\simeq2.50\times10^{5}$ yr. 
Although this estimate assumes a constant shock velocity, it should still provide a reasonable characteristic age: the hot gas cooling time is much longer than $t_{\rm dyn}$, and the Mach number is unlikely to have varied by more than a factor of a few during the current phase.

\begin{figure}
\centering
\includegraphics[width=\columnwidth]{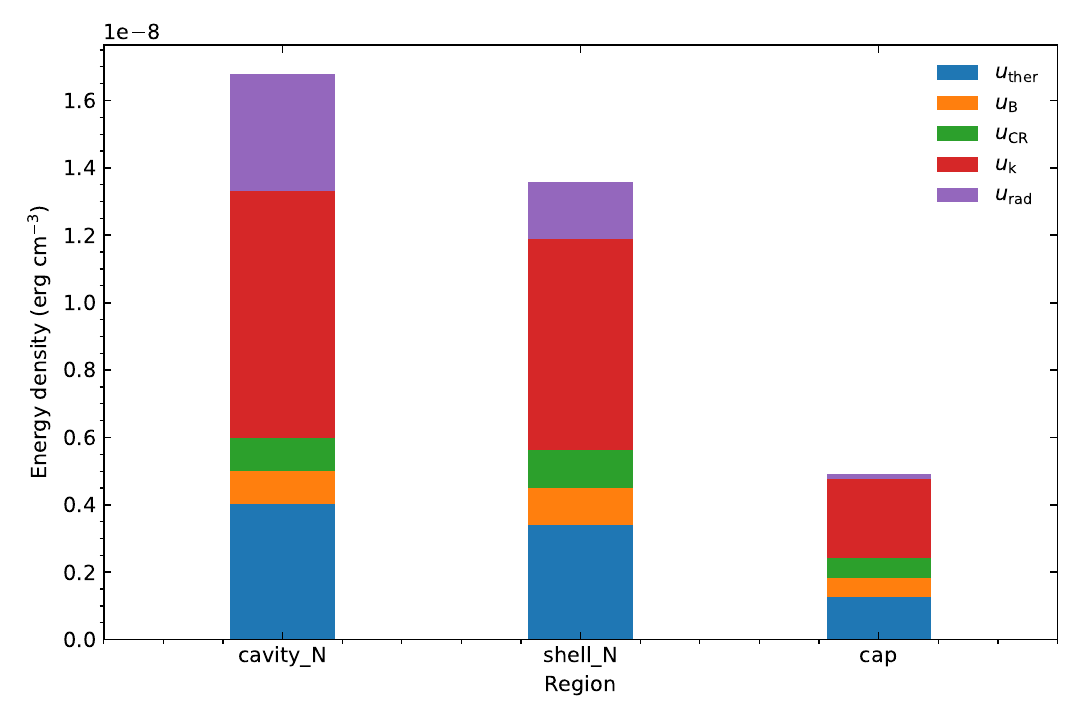}
\caption{
Energy density budget of the northwest nuclear bubble in NGC~4438, shown separately for the cavity, shell, and cap regions. 
For each region, the stacked bars represent the contributions from the thermal energy of the X-ray--emitting gas ($u_{\rm ther}$), magnetic energy ($u_B$), cosmic-ray energy ($u_{CR}$), bulk kinetic energy ($u_{\rm kin}$), and cumulative radiative losses ($u_{\rm rad}$). 
All energy densities are computed using the same geometric assumptions described in Section~\ref{sec:energy_budget_baby}. 
The comparison illustrates how different energy channels are partitioned spatially within the jet-driven bubble, with the shell dominated by kinetic energy and the cavity and cap exhibiting substantial contributions from thermal and magnetic components.
}
\label{fig:energy_density_baby}
\end{figure}

Using this timescale, the cumulative radiative energy loss of the northern bubble
is estimated as
\begin{equation}
E_{\rm rad,N} = (L_X + L_{\mathrm{H\alpha+[N\,II]}})\, t_{\rm dyn}
\simeq 3.30\times10^{53}\ {\rm erg},
\end{equation}
while the bulk kinetic energy is
\begin{equation}
E_{\rm kin,N} = \frac{1}{2} M v_{\rm sh,N}^2
\simeq 1.74\times10^{54}\ {\rm erg},
\end{equation}
where the hot-gas mass, computed from the measured electron density and the adopted volume of each subregion, is $M_{\rm hot,N}\simeq1.4\times10^{5}\ M_\odot$. 
The thermal and magnetic/cosmic-ray energies are obtained by summing the corresponding subregions listed in Table~\ref{tab:fitting_small_bubble}. 
We then define the total energy budget of the bubble as
\begin{equation}
E_{\rm tot}=E_{\rm ther}+E_{CR}+E_B+E_{\rm rad}+E_{\rm kin},
\end{equation}
which gives
$E_{\rm tot,N}\gtrsim3.1\times10^{54}$ erg. The corresponding characteristic power is
\begin{equation}
P_{\rm tot,N}=\frac{E_{\rm tot,N}}{t_{\rm dyn}}\simeq3.9\times10^{41} {\rm erg s}^{-1}.
\end{equation}
The corresponding energy densities are shown in
Figure~\ref{fig:energy_density_baby}. 
Despite the systematic uncertainty in the shock velocity and dynamical time, which may approach a factor of two, the overall energy partition is robust: bulk kinetic energy remains the dominant reservoir, while the thermal and radiative terms are comparable, and the magnetic/cosmic-ray component remains non-negligible. 
Consistent with the plasma $\beta$ values (see Table~\ref{tab:fitting_small_bubble}), the thermal and magnetic pressures are also broadly comparable, suggesting that both hot gas and magnetic fields contribute to the bubble expansion even though the system as a whole is still in a dynamically active, shock-dominated phase in which a substantial fraction of the injected jet energy has not yet been fully thermalized.

We also derive a simplified energetic estimate for the southeast nuclear bubble.
Because of the much weaker H$\alpha$ and X-ray emission, likely compounded by stronger dust extinction, a full radiative budget is not feasible. 
We therefore restrict the southern analysis to the thermal, magnetic/cosmic-ray, and kinetic components, adopting the same geometric assumptions as for the northern side.
Assuming that the northern and southern bubbles were launched quasi-simultaneously by the same AGN episode, we adopt the same characteristic dynamical time, $t_{\rm dyn}=2.50\times10^{5}$ yr. 
The projected distance to the southern shell of $R_{\rm proj,S}=9.8\arcsec$ implies $v_{\rm sh,S}\simeq2.7\times10^{3}$ km s$^{-1}$. 
Using X-ray-derived densities and the same geometric assumptions, we estimate a hot-gas mass of $M_{\rm hot,S}\simeq1.3\times10^{5}\ M_\odot$, corresponding to a bulk kinetic energy of $E_{\rm kin,S}\simeq1.0\times10^{55}$ erg. 
The thermal and magnetic/cosmic-ray energies taken from the summed shell$_S$ and cavity$_S$ entries in Table~\ref{tab:fitting_small_bubble} are $E_{\rm ther,S}\simeq5.6\times10^{53}$ erg and $E_{CR,S}=E_{B,S}\simeq2.8\times10^{53}$ erg, giving a lower limit on the total energy of $E_{\rm tot,S}\gtrsim1.05\times10^{55}$ erg and a corresponding lower limit on the characteristic power of $P_{\rm bub,S}\gtrsim1.33\times10^{42}$ erg s$^{-1}$.

The southern bubble, therefore, appears to retain a larger fraction of its energy in bulk kinetic form and to have propagated farther from the nucleus at a higher velocity over a similar dynamical timescale. 
This asymmetry is naturally explained if the southern jet is expanding into a lower-density environment, allowing it to remain faster and less thermalized, while stronger interaction with denser gas on the northern side enhances shock heating and deceleration. 
Despite these differences, the total energy content of the two nuclear bubbles is broadly comparable within uncertainties, consistent with a common jet-driven origin.

The $4pV$ cavity-enthalpy estimate has been established primarily through studies of evolved bubbles and radio-filled cavities on kiloparsec and larger scales \citep[e.g.,][]{Birzan2004,Rafferty2006,McNamaraNulsen2007,Fabian2012}. In such systems, the cavity is generally assumed to evolve slowly and to remain approximately in pressure balance with the surrounding hot atmosphere, so that its enthalpy can be written as
\begin{equation}
E_{\rm cav}=\frac{\gamma}{\gamma-1}pV\simeq4pV
\end{equation}
for a relativistic equation of state with $\gamma=4/3$ \citep{Birzan2004,Allen2006}. These assumptions are unlikely to hold for the compact nuclear bubbles in NGC~4438. Their sub-kpc sizes, strong shocks, and rapid expansion indicate that they have not yet reached dynamical or pressure equilibrium. The $4pV$ estimate therefore should not be expected to provide an adequate measure of their total injected energy.

The numerical comparison confirms this breakdown. Using the X-ray-derived thermal pressures and the same volumes adopted above, we obtain $E_{4pV,{\rm N}}\simeq1.4\times10^{54}$ erg and $P_{4pV,{\rm N}}\simeq1.8\times10^{41}$ erg s$^{-1}$ for the NW nuclear bubble, and $E_{4pV,{\rm S}}\simeq1.9\times10^{54}$ erg and $P_{4pV,{\rm S}}\simeq2.4\times10^{41}$ erg s$^{-1}$ for the SE nuclear bubble. These values are lower than the powers inferred from the resolved energy budgets, $P_{\rm tot,N}\simeq3.9\times10^{41}$ erg s$^{-1}$ and $P_{\rm tot,S}\gtrsim1.33\times10^{42}$ erg s$^{-1}$, by factors of approximately 2 and 5, respectively. The discrepancy is expected for rapidly expanding bubbles because $4pV$ describes the enthalpy of a pressure-supported cavity but does not account for the substantial bulk kinetic energy retained by the shocked gas or for energy already dissipated through radiation. Thus, applying the $4pV$ formalism to these nuclear bubbles systematically underestimates their energetics.

This limitation also applies to the empirical radio--jet power relation established by \citet{MerloniHeinz2007}. That relation was calibrated using a sample of AGN-bubble systems with kiloparsec-scale and larger X-ray cavities, for which the jet powers were themselves inferred from $4pV$ cavity enthalpies divided by characteristic cavity ages. It therefore inherits the same physical assumptions as the cavity-enthalpy method, in particular that the observed structures are evolved, approximately pressure-supported cavities whose energetics can be represented by $4pV$. These assumptions are not satisfied by the compact, shock-dominated nuclear bubbles in NGC~4438. Nevertheless, applying the \citet{MerloniHeinz2007} relation to the nuclear radio luminosity led \citet{PuigSubira2026} to infer a jet kinetic power of $\sim5\times10^{44}$ erg s$^{-1}$, more than three orders of magnitude above the powers obtained from our resolved energy budgets. This discrepancy is therefore not simply a difference between two independent power estimates; rather, it reflects the extrapolation of a relation calibrated from large, evolved $4pV$ cavities into a much smaller, dynamically unrelaxed regime. The \citet{MerloniHeinz2007} relation consequently cannot be applied reliably to the nuclear bubbles in NGC~4438.

\subsubsection{Comparing the Nuclear Bubbles with the Galaxy-scale Outflow}
\label{sec:large_bubble_timescale}

Having quantified the resolved energy budget of the nuclear bubbles, we now turn to the galaxy-scale outflow.
Our goal is to examine whether the energetics and timescales of the compact nuclear bubbles are broadly consistent with those required to produce the much larger galaxy-scale outflow, and thus to assess whether the two structures may be physically connected as different evolutionary manifestations of the same feedback process.

As a first step, we derive a simple order-of-magnitude estimate for the galaxy-scale outflow based on its present-day enthalpy and a sound-crossing timescale. 
Using the thermal pressures listed in Table~\ref{tab:fitting_large_bubble} and considering only the cavity, shell, and cap regions, we estimate the enthalpy of the present cavity using Eq.~(9), where the volume of each subregion is approximated as the projected area times a characteristic line-of-sight depth of 8.4 kpc. 
This gives a total enthalpy of $E_{\rm cav,large}\simeq4.6\times10^{56}$ erg. Similarly, if the observed bubble size is $R_{\rm obs}\simeq8.4$~kpc and the surrounding medium has $kT_0\simeq0.62$~keV (Table~\ref{tab:fitting_large_bubble}), the ambient sound speed is,
\begin{equation}
c_{s,0}=
\left(\frac{\gamma k_{\rm B}T_0}{\mu m_p}\right)^{1/2}
\simeq 4.05\times10^2~{\rm km\,s^{-1}},
\end{equation}
implying a simple sound-crossing time of
\begin{equation}
t_{\rm dyn}\equiv \frac{R_{\rm obs}}{c_{s,0}}
\simeq 2.0\times10^7~{\rm yr}.
\end{equation}
These estimates are useful as a first guide, but they implicitly assume that the bubble evolution can be characterized by its present-day pressure and by expansion at roughly the ambient sound speed throughout its history.

A more physical estimate should account for the fact that the galaxy-scale outflow is still mildly overpressured and is likely already in a late, pressure-regulated stage of evolution. 
For this purpose, we adopt the analytic framework of \citet{Li2026_model}, who modeled the evolution of a feedback-driven bubble that expanded into a hot ambient medium filled with volume. 
In that framework, the bubble evolves from a Sedov--Taylor-like stage into a pressure-modified stage, and the end of the strong-shock phase is determined by pressure balance and/or the transition to transonic expansion rather than by rapid radiative cooling. 
This description is well-suited for the galaxy-scale outflow in NGC~4438, whose internal pressure is only moderately higher than that of the surrounding hot gas.

The galaxy-scale outflow in NGC~4438 appears to be well suited to this late-stage description.
As listed in Table~\ref{tab:fitting_large_bubble}, its internal pressure, defined as
\begin{equation}
p_{\rm in}=p_{\rm th}+p_B\simeq 0.0068~{\rm keV\,cm^{-3}},
\end{equation}
while the ambient thermal pressure is $p_{out}\simeq 0.0043~{\rm keV\,cm^{-3}}$. 
The resulting ratio, $p_{\rm in}/p_{\rm out}\simeq1.58$, indicates that the bubble is only slightly overpressured, suggesting that it is already close to the pressure-balanced transonic regime identified by \citet{Li2026_model}, i.e., their Stage III. 
In this stage, the integrity and continued expansion of the galaxy-scale outflow are maintained by the combined action of thermal and magnetic pressure. 
Consistent with this picture, the plasma $\beta$ of the galaxy-scale outflow is slightly lower than that of the nuclear bubbles (Tables~\ref{tab:fitting_small_bubble} and \ref{tab:fitting_large_bubble}), implying a modestly greater relative role of magnetic pressure on larger scales.

We therefore use the environmental parameters measured for the surrounding filamentary medium in Table~\ref{tab:fitting_large_bubble}, namely $kT_0\simeq0.62$~keV and $n_{e,0}\simeq3.6\times10^{-3}~{\rm cm^{-3}}$, together with the observed bubble radius $R_{\rm obs}\simeq8.4$~kpc. 
Following \citet{Li2026_model}, the characteristic pressure-balance radius is given by their Eq.~(25),
\begin{equation}
R_p \simeq 1.1\,E_{51}^{1/3} n_{-3}^{-1/3} T_6^{-1/3}~{\rm kpc},
\end{equation}
where $E_{51}\equiv E_0/(10^{51}~{\rm erg})$, $n_{-3}\equiv n_0/(10^{-3}~{\rm cm^{-3}})$, and $T_6\equiv T_0/(10^6~{\rm K})$. 
Combining their Eqs.~(23)--(25), or equivalently using the Sedov--Taylor pressure scaling $p_b\propto R^{-3}$ from their Eq.~(24), the current mild overpressure implies
\begin{equation}
\frac{R_p}{R_{\rm obs}}=
\left(\frac{p_{\rm in}}{p_{\rm out}}\right)^{1/3}.
\end{equation}
Substituting the measured pressure ratio gives $R_p\simeq9.8$~kpc, which we take as a good approximation to the detectable-shell radius, $R_{\rm det}$, defined by \citet{Li2026_model} in their Eq.~(83) as $R_{\rm det}\equiv R(t_{\rm det})$ with $t_{\rm det}\equiv\max(t_p,t_M)$.

Using this radius and the ambient temperature and density in the above expression, we infer an equivalent injected energy of $E_0\simeq3.5\times10^{55}$~erg.
In the same framework, the pressure-balance timescale is given by their Eq.~(26),
\begin{equation}
t_p \simeq 2.6\times10^6\,E_{51}^{1/3} n_{-3}^{-1/3} T_6^{-5/6}~{\rm yr},
\end{equation}
while the transonic time scale is similarly given by their Eq.~(30),
\begin{equation}
t_M \simeq 3.0\times10^6\,E_{51}^{1/3} n_{-3}^{-1/3} T_6^{-5/6}~{\rm yr}.
\end{equation}
Substituting the measured parameters yields $t_p\simeq8.6$~Myr and $t_M\simeq10.0$~Myr, and therefore $t_{\rm det}\simeq10$~Myr.

The bubble's current age can also be estimated from its overpressure. 
Here we extrapolate the Sedov--Taylor pressure scaling implied by their Eqs.~(24) and (27), i.e. $p_b\propto t^{-6/5}$, to the current mildly overpressured state. This estimate is approximate, since the bubble is likely already in the pressure-modified Stage III, where the classical Sedov--Taylor similarity solution no longer strictly applies. Then we have
\begin{equation}
\frac{t_p}{t_{\rm current}}=
\left(\frac{p_{\rm in}}{p_{\rm out}}\right)^{5/6},
\end{equation}
which gives $t_{\rm current}\simeq5.9$~Myr.
Thus, under the \citet{Li2026_model} interpretation, the galaxy-scale outflow in NGC~4438 is already in a relatively late evolutionary stage: its present radius of 8.4 kpc is close to the characteristic maximum coherent size of $\sim9.8$ kpc, while its current age is likely already a substantial fraction of the $\sim10$ Myr timescale over which the outflow remains identifiable as a coherent shell/cavity structure.
The inferred energy and timescale of the galaxy-scale outflow allow us to assess whether it can be physically linked to the present-day nuclear bubbles.
The derived values, $E_0\simeq3.5\times10^{55}$~erg and $t_{\rm current}\simeq5.9$~Myr, imply a characteristic mean power requirement of $\dot E_{\rm req}\simeq1.9\times10^{41}$~erg~s$^{-1}$ to build up the observed large-scale outflow.

A simple estimate of the mechanical power supplied by star formation is
\begin{equation}
\dot E_{\rm SF} \simeq 3.2\times10^{41}
\left(\frac{{\rm SFR}}{1~M_\odot~{\rm yr}^{-1}}\right)
{\rm erg~s^{-1}},
\end{equation}
assuming one core-collapse supernova releases $\sim10^{51}$~erg per $\sim100~M_\odot$ of newly formed stars.
Matching $\dot E_{\rm req}$ would therefore require ${\rm SFR}\sim0.6~M_\odot~{\rm yr}^{-1}$, before accounting for uncertain coupling efficiencies.
This is comparable to the total star formation rate of the entire galaxy ($\sim0.5~M_\odot~{\rm yr}^{-1}$; \citealt{Mahajan2019}), and substantially higher than the nuclear star formation rate ($\sim0.05$--$0.1~M_\odot~{\rm yr}^{-1}$; \citealt{Hota2007}).
It is therefore unlikely that nuclear star formation alone can power the galaxy-scale outflow.

In contrast, the NW nuclear bubble has $\dot P_{\rm tot,N}\simeq3.9\times10^{41}$~erg~s$^{-1}$.
This exceeds the mean power required for the galaxy-scale outflow by a factor of $\sim2$. 
If the AGN injects energy at a level comparable to that traced by the present-day NW nuclear bubble, the duty cycle required to sustain large-scale outflow would be $\dot E_{\rm req}/\dot E_{\rm nuc,N}\sim0.5$.
An activity duty cycle of the order of fifty percent over several Myr would therefore suffice to produce the observed galaxy-scale outflow.

This duty cycle is not unusually high in the context of mechanical AGN feedback. 
X-ray cavity studies in groups and clusters typically infer duty cycles of order $\sim60\%$, increasing to $\gtrsim80\%$ in systems with short central cooling times \citep{Birzan2012,Panagoulia2014,Fabian2012}.
Although these measurements refer to hot halos that are more massive and dynamically relaxed than those of NGC~4438, they indicate that a duty cycle of $\sim50\%$ is entirely plausible for sustained jet activity.

In NGC~4438, the disturbed environment may further facilitate such prolonged feedback. 
The galaxy is undergoing strong tidal interaction and ram-pressure stripping, which have significantly redistributed its multiphase gas \citep{Hota2007,Vollmer2009}. 
These conditions may both enhance gas inflow toward the nucleus and reduce the effective confinement of large-scale outflows within a partially stripped interstellar medium.
The required duty cycle, therefore, does not appear excessive and may be naturally linked to the interaction-driven state of the system.

\section{CONCLUSIONS}\label{sec:conclusion}

We present a spatially resolved multi-wavelength analysis of AGN jet-driven feedback in NGC~4438, combining deep \textit{Chandra} X-ray data with optical H$\alpha$ imaging and VLA radio continuum observations.
By jointly examining the compact nuclear bubbles and the surrounding $\sim10$-kpc galaxy-scale outflow, we investigate how jet energy is converted into kinetic, thermal, magnetic, and radiative components across spatial scales.
Our main conclusions are summarized as follows.

\begin{enumerate}

\item
The central region of NGC~4438 hosts two oppositely directed sub-kiloparsec jet-driven bubbles, embedded within a much larger-scale outflow structure.
In particular, the northern bubble exhibits a radio-bright cap ahead of the shell, which we interpret as a magnetic draping feature and which highlights strong spatial variations in the non-thermal components.

\item
Joint radio and X-ray spectral modeling demonstrates that the non-thermal emission associated with the nuclear bubbles cannot be explained by a single electron population.
Instead, extrapolating the radio synchrotron component to the X-ray band with a single power law overpredicts the observed hard X-ray emission by nearly eight orders of magnitude, while the opposite spatial trends of the radio and X-ray spectral indices further imply at least two distinct electron populations and two acceleration channels.
The GeV electron population is most naturally explained by shock acceleration at the bubble rim, whereas the origin of the TeV electrons is more likely tied to AGN-related processes operating within the jet or bubble interior.

\item
A spatially resolved energy census of the nuclear bubbles shows that bulk kinetic energy remains the dominant energy reservoir of the expanding bubble at the current epoch.
In the northern bubble, about half of the injected energy has already been converted into thermal energy of the hot gas, CR, magnetic energy, and radiative losses, indicating efficient energy dissipation.
In contrast, the southern bubble appears to retain a larger fraction of its energy in kinetic form, possibly due to propagation into a lower-density environment where shock heating and radiative losses are less effective.

\item
By comparing different approaches to estimating jet energetics, we show that commonly used methods must be applied with caution on sub-kiloparsec scales.
For the compact nuclear bubbles, the cavity enthalpy estimate ($4pV$) systematically underestimates the total injected energy because the system has not yet entered a quasi-adiabatic expansion phase and retains a large kinetic energy component.
Conversely, radio-based empirical relations \citep{MerloniHeinz2007} calibrated on kpc scale cavities substantially overestimate the jet power when applied to such young, compact bubbles.

\item
Extending the energy analysis to the galaxy-scale outflow reveals that it remains supported by both thermal and magnetic pressure, with a slightly lower plasma $\beta$ than in the nuclear bubbles.
Its energetics and timescale are broadly consistent with sustained AGN jet activity at a level comparable to that traced by the present-day northwest nuclear bubble.
This outflow can be powered by AGN jet activity with a plausible duty cycle of order 50\% over several Myr, whereas nuclear star formation alone is unlikely to account for the required energetics.

\end{enumerate}

Overall, our results demonstrate that even a low-luminosity LINER-type AGN such as NGC~4438 can drive energetically significant multi-phase outflows.
By resolving the energy budget and non-thermal processes from $\sim$200~pc to 10-kpc scales within the same system, this study highlights the importance of spatially resolved, multi-wavelength observations for understanding how the AGN jet feedback evolves and couples to the surrounding medium.

\begin{acknowledgements}

This work was supported in part by the National Natural Science Foundation of China (NSFC; grants 12321003 and 12273111), the China Manned Space Program (grants CMS-CSST-2025-A04 and CMS-CSST-2025-A10), and the Jiangsu Innovation and Entrepreneurship Talent Team Program (grant JSSCTD202436). 
JX and GL additionally acknowledge support from the Ministry of Science and Technology of China (grant 2023YFA1608100), the China Manned Space Project (grant CMS-CSST-2025-A08), and the National Natural Science Foundation of China (grant 12273036). 
Y.Y. acknowledges support from the National Natural Science Foundation of China through the grant 12203098. 
We are grateful to J. D. P. Kenney for kindly providing the KPNO 4 m telescope H$\alpha$ data used in this work. All the {\it HST} data used in this paper can be found in MAST: \dataset[10.17909/0qkg-vg37]{http://dx.doi.org/10.17909/0qkg-vg37}. We thank the anonymous referee for the constructive comments that significantly improved the clarity and precision of both the figures and the presentation of this work.

The authors acknowledge the use of ChatGPT (OpenAI, San Francisco, USA) for assistance with coding and for improving the clarity and grammar of the manuscript, and the Cursor editor for assistance with code development. 
All scientific ideas, analyses, interpretations, and conclusions presented in this work were conceived and verified by the authors, who take full responsibility for the content.

\end{acknowledgements}

\bibliographystyle{aasjournalv7}
\bibliography{4438_paper}

@ARTICLE{Wang2021M83,
       author = {{Wang}, Q. Daniel and {Zeng}, Yuxuan and {Bogd{\'a}n}, {\'A}kos and {Ji}, Li},
        title = "{Deep Chandra observations of diffuse hot plasma in M83}",
      journal = {\mnras},
         year = 2021,
        month = dec,
       volume = {508},
       number = {4},
        pages = {6155-6175},
          doi = {10.1093/mnras/stab2997},
archivePrefix = {arXiv},
       eprint = {2110.06995},
 primaryClass = {astro-ph.GA},
       adsurl = {https://ui.adsabs.harvard.edu/abs/2021MNRAS.508.6155W}
}

@ARTICLE{Suzuki2021,
       author = {{Suzuki}, H. and {Plucinsky}, P.~P. and {Gaetz}, T.~J. and {Bamba}, A.},
        title = "{Spatial and temporal variations of the Chandra ACIS particle-induced background and development of a spectral-model generation tool}",
      journal = {\aap},
         year = 2021,
        month = nov,
       volume = {655},
          eid = {A116},
        pages = {A116},
          doi = {10.1051/0004-6361/202141458},
archivePrefix = {arXiv},
       eprint = {2108.11234},
 primaryClass = {astro-ph.HE},
       adsurl = {https://ui.adsabs.harvard.edu/abs/2021A&A...655A.116S}
}

@ARTICLE{LuanWang2025,
       author = {{Luan}, Luan and {Wang}, Q. Daniel},
        title = "{Diffuse X-Ray Emission in M51: A Hierarchical Bayesian Spatially Resolved Spectral Analysis}",
      journal = {\apj},
         year = 2025,
        month = jun,
       volume = {986},
       number = {2},
          eid = {167},
        pages = {167},
          doi = {10.3847/1538-4357/adcf16},
archivePrefix = {arXiv},
       eprint = {2504.15641},
 primaryClass = {astro-ph.HE},
       adsurl = {https://ui.adsabs.harvard.edu/abs/2025ApJ...986..167L}
}

@ARTICLE{Kenney2008,
       author = {{Kenney}, Jeffrey D.~P. and {Tal}, Tomer and {Crowl}, Hugh H. and {Feldmeier}, John and {Jacoby}, George H.},
        title = "{A Spectacular H{\ensuremath{\alpha}} Complex in Virgo: Evidence for a Collision between M86 and NGC 4438 and Implications for the Collisional ISM Heating of Ellipticals}",
      journal = {\apjl},
         year = 2008,
        month = nov,
       volume = {687},
       number = {2},
        pages = {L69},
          doi = {10.1086/593300},
archivePrefix = {arXiv},
       eprint = {0810.0711},
 primaryClass = {astro-ph},
       adsurl = {https://ui.adsabs.harvard.edu/abs/2008ApJ...687L..69K}
}

@ARTICLE{Hermosa2024,
       author = {{Hermosa Mu{\~n}oz}, L. and {Cazzoli}, S. and {M{\'a}rquez}, I. and {Masegosa}, J. and {Chamorro-Cazorla}, M. and {Gil de Paz}, A. and {Castillo-Morales}, {\'A}. and {Gallego}, J. and {Carrasco}, E. and {Iglesias-P{\'a}ramo}, J. and {Garc{\'\i}a-Vargas}, M.~L. and {G{\'o}mez-{\'A}lvarez}, P. and {Pascual}, S. and {P{\'e}rez-Calpena}, A. and {Cardiel}, N.},
        title = "{The MEGARA view of outflows in LINERs}",
      journal = {\aap},
         year = 2024,
        month = mar,
       volume = {683},
          eid = {A43},
        pages = {A43},
          doi = {10.1051/0004-6361/202347675},
archivePrefix = {arXiv},
       eprint = {2311.12552},
 primaryClass = {astro-ph.GA},
       adsurl = {https://ui.adsabs.harvard.edu/abs/2024A&A...683A..43H}
}

@ARTICLE{Li2022,
       author = {{Li}, Jiang-Tao and {Wang}, Q. Daniel and {Wiegert}, Theresa and {Bregman}, Joel N. and {Beck}, Rainer and {Damas-Segovia}, Ancor and {Irwin}, Judith A. and {Ji}, Li and {Stein}, Yelena and {Sun}, Wei and {Yang}, Yang},
        title = "{CHANG-ES XXIX: the sub-kpc nuclear bubble of NGC 4438}",
      journal = {\mnras},
         year = 2022,
        month = sep,
       volume = {515},
       number = {2},
        pages = {2483-2495},
          doi = {10.1093/mnras/stac837},
archivePrefix = {arXiv},
       eprint = {2205.12343},
 primaryClass = {astro-ph.HE},
       adsurl = {https://ui.adsabs.harvard.edu/abs/2022MNRAS.515.2483L}
}

@ARTICLE{Li2024,
       author = {{Li}, Jiang-Tao and {Sun}, Wei and {Ji}, Li and {Yang}, Yang},
        title = "{Pressure Balance and Energy Budget of the Nuclear Superbubble of NGC 3079}",
      journal = {\apj},
         year = 2024,
        month = may,
       volume = {966},
       number = {2},
          eid = {239},
        pages = {239},
          doi = {10.3847/1538-4357/ad3af2},
archivePrefix = {arXiv},
       eprint = {2404.03879},
 primaryClass = {astro-ph.GA},
       adsurl = {https://ui.adsabs.harvard.edu/abs/2024ApJ...966..239L}
}

@ARTICLE{PuigSubira2026,
       author = {{Puig-Subir{\`a}}, M. and {Mold{\'o}n}, J. and {M{\'a}rquez}, I. and {Masegosa}, J. and {Gonz{\'a}lez-Mart{\'\i}n}, O. and {Hermosa Mu{\~n}oz}, L. and {Cazzoli}, S. and {Williams-Baldwin}, D.},
        title = "{A multi-wavelength approach of AGN feedback in LINERs: The case of NGC 4438}",
      journal = {arXiv e-prints},
         year = 2026,
        month = jan,
          eid = {arXiv:2601.15991},
        pages = {arXiv:2601.15991},
          doi = {10.48550/arXiv.2601.15991},
archivePrefix = {arXiv},
       eprint = {2601.15991},
 primaryClass = {astro-ph.GA},
       adsurl = {https://ui.adsabs.harvard.edu/abs/2026arXiv260115991P}
}

@ARTICLE{MerloniHeinz2007,
       author = {{Merloni}, Andrea and {Heinz}, Sebastian},
        title = "{Measuring the kinetic power of active galactic nuclei in the radio mode}",
      journal = {\mnras},
         year = 2007,
        month = oct,
       volume = {381},
       number = {2},
        pages = {589-601},
          doi = {10.1111/j.1365-2966.2007.12253.x},
archivePrefix = {arXiv},
       eprint = {0707.3356},
 primaryClass = {astro-ph},
       adsurl = {https://ui.adsabs.harvard.edu/abs/2007MNRAS.381..589M}
}

@ARTICLE{Li2019,
       author = {{Li}, Jiang-Tao and {Hodges-Kluck}, Edmund and {Stein}, Yelena and {Bregman}, Joel N. and {Irwin}, Judith A. and {Dettmar}, Ralf-J{\"u}rgen},
        title = "{Detection of Nonthermal Hard X-Ray Emission from the {\textquotedblleft}Fermi Bubble{\textquotedblright} in an External Galaxy}",
      journal = {\apj},
         year = 2019,
        month = mar,
       volume = {873},
       number = {1},
          eid = {27},
        pages = {27},
          doi = {10.3847/1538-4357/ab010a},
archivePrefix = {arXiv},
       eprint = {1901.10536},
 primaryClass = {astro-ph.HE},
       adsurl = {https://ui.adsabs.harvard.edu/abs/2019ApJ...873...27L}
}

@ARTICLE{SironiSpitkovsky2014,
       author = {{Sironi}, Lorenzo and {Spitkovsky}, Anatoly},
        title = "{Relativistic Reconnection: An Efficient Source of Non-thermal Particles}",
      journal = {\apjl},
         year = 2014,
        month = mar,
       volume = {783},
       number = {1},
          eid = {L21},
        pages = {L21},
          doi = {10.1088/2041-8205/783/1/L21},
archivePrefix = {arXiv},
       eprint = {1401.5471},
 primaryClass = {astro-ph.HE},
       adsurl = {https://ui.adsabs.harvard.edu/abs/2014ApJ...783L..21S}
}

@ARTICLE{Guo2014,
       author = {{Guo}, Fan and {Li}, Hui and {Daughton}, William and {Liu}, Yi-Hsin},
        title = "{Formation of Hard Power Laws in the Energetic Particle Spectra Resulting from Relativistic Magnetic Reconnection}",
      journal = {\prl},
         year = 2014,
        month = oct,
       volume = {113},
       number = {15},
          eid = {155005},
        pages = {155005},
          doi = {10.1103/PhysRevLett.113.155005},
archivePrefix = {arXiv},
       eprint = {1405.4040},
 primaryClass = {astro-ph.HE},
       adsurl = {https://ui.adsabs.harvard.edu/abs/2014PhRvL.113o5005G}
}

@ARTICLE{Giannios2013,
       author = {{Giannios}, Dimitrios},
        title = "{Reconnection-driven plasmoids in blazars: fast flares on a slow envelope}",
      journal = {\mnras},
         year = 2013,
        month = may,
       volume = {431},
       number = {1},
        pages = {355-363},
          doi = {10.1093/mnras/stt167},
archivePrefix = {arXiv},
       eprint = {1211.0296},
 primaryClass = {astro-ph.HE},
       adsurl = {https://ui.adsabs.harvard.edu/abs/2013MNRAS.431..355G}
}

@ARTICLE{Levinson2000,
       author = {{Levinson}, Amir},
        title = "{Particle Acceleration and Curvature TeV Emission by Rotating, Supermassive Black Holes}",
      journal = {\prl},
         year = 2000,
        month = jul,
       volume = {85},
       number = {5},
        pages = {912-915},
          doi = {10.1103/PhysRevLett.85.912},
       adsurl = {https://ui.adsabs.harvard.edu/abs/2000PhRvL..85..912L}
}

@article{RiegerAharonian2008,
       author = {{Rieger}, F.~M. and {Aharonian}, F.~A.},
        title = "{Variable VHE gamma-ray emission from non-blazar AGNs}",
      journal = {\aap},
         year = 2008,
        month = feb,
       volume = {479},
       number = {1},
        pages = {L5-L8},
          doi = {10.1051/0004-6361:20078706},
archivePrefix = {arXiv},
       eprint = {0712.2902},
 primaryClass = {astro-ph},
       adsurl = {https://ui.adsabs.harvard.edu/abs/2008A&A...479L...5R}
}

@article{Spada2001,
       author = {{Spada}, M. and {Lazzati}, D. and {Ghisellini}, G. and {Celotti}, A.},
        title = "{Internal shocks in the jets of blazars}",
      journal = {\memsai},
         year = 2001,
        month = jan,
       volume = {72},
        pages = {157-159},
          doi = {10.48550/arXiv.astro-ph/0006372},
archivePrefix = {arXiv},
       eprint = {astro-ph/0006372},
 primaryClass = {astro-ph},
       adsurl = {https://ui.adsabs.harvard.edu/abs/2001MmSAI..72..157S}
}

@article{BlandfordKonigl1979,
       author = {{Blandford}, R.~D. and {K{\"o}nigl}, A.},
        title = "{Relativistic jets as compact radio sources.}",
      journal = {\apj},
         year = 1979,
        month = aug,
       volume = {232},
        pages = {34-48},
          doi = {10.1086/157262},
       adsurl = {https://ui.adsabs.harvard.edu/abs/1979ApJ...232...34B}
}

@article{McNamaraNulsen2007,
       author = {{McNamara}, B.~R. and {Nulsen}, P.~E.~J.},
        title = "{Heating Hot Atmospheres with Active Galactic Nuclei}",
      journal = {\araa},
         year = 2007,
        month = sep,
       volume = {45},
       number = {1},
        pages = {117-175},
          doi = {10.1146/annurev.astro.45.051806.110625},
archivePrefix = {arXiv},
       eprint = {0709.2152},
 primaryClass = {astro-ph},
       adsurl = {https://ui.adsabs.harvard.edu/abs/2007ARA&A..45..117M}
}

@article{Fabian2012,
       author = {{Fabian}, A.~C.},
        title = "{Observational Evidence of Active Galactic Nuclei Feedback}",
      journal = {\araa},
         year = 2012,
        month = sep,
       volume = {50},
        pages = {455-489},
          doi = {10.1146/annurev-astro-081811-125521},
archivePrefix = {arXiv},
       eprint = {1204.4114},
 primaryClass = {astro-ph.CO},
       adsurl = {https://ui.adsabs.harvard.edu/abs/2012ARA&A..50..455F}
}

@article{HeckmanBest2014,
       author = {{Heckman}, Timothy M. and {Best}, Philip N.},
        title = "{The Coevolution of Galaxies and Supermassive Black Holes: Insights from Surveys of the Contemporary Universe}",
      journal = {\araa},
         year = 2014,
        month = aug,
       volume = {52},
        pages = {589-660},
          doi = {10.1146/annurev-astro-081913-035722},
archivePrefix = {arXiv},
       eprint = {1403.4620},
 primaryClass = {astro-ph.GA},
       adsurl = {https://ui.adsabs.harvard.edu/abs/2014ARA&A..52..589H}
}

@article{BegelmanCioffi1989,
       author = {{Begelman}, Mitchell C. and {Cioffi}, Denis F.},
        title = "{Overpressured Cocoons in Extragalactic Radio Sources}",
      journal = {\apjl},
         year = 1989,
        month = oct,
       volume = {345},
        pages = {L21},
          doi = {10.1086/185542},
       adsurl = {https://ui.adsabs.harvard.edu/abs/1989ApJ...345L..21B}
}

@article{Churazov2002,
       author = {{Churazov}, E. and {Br{\"u}ggen}, M. and {Kaiser}, C.~R. and {B{\"o}hringer}, H. and {Forman}, W.},
        title = "{Evolution of Buoyant Bubbles in M87}",
      journal = {\apj},
         year = 2001,
        month = jun,
       volume = {554},
       number = {1},
        pages = {261-273},
          doi = {10.1086/321357},
archivePrefix = {arXiv},
       eprint = {astro-ph/0008215},
 primaryClass = {astro-ph},
       adsurl = {https://ui.adsabs.harvard.edu/abs/2001ApJ...554..261C}
}

@article{Croston2005,
       author = {{Croston}, J.~H. and {Hardcastle}, M.~J. and {Harris}, D.~E. and {Belsole}, E. and {Birkinshaw}, M. and {Worrall}, D.~M.},
        title = "{An X-Ray Study of Magnetic Field Strengths and Particle Content in the Lobes of FR II Radio Sources}",
      journal = {\apj},
         year = 2005,
        month = jun,
       volume = {626},
       number = {2},
        pages = {733-747},
          doi = {10.1086/430170},
archivePrefix = {arXiv},
       eprint = {astro-ph/0503203},
 primaryClass = {astro-ph},
       adsurl = {https://ui.adsabs.harvard.edu/abs/2005ApJ...626..733C}
}

@ARTICLE{HardcastleCroston2020,
       author = {{Hardcastle}, M.~J. and {Croston}, J.~H.},
        title = "{Radio galaxies and feedback from AGN jets}",
      journal = {\nar},
         year = 2020,
        month = jun,
       volume = {88},
          eid = {101539},
        pages = {101539},
          doi = {10.1016/j.newar.2020.101539},
archivePrefix = {arXiv},
       eprint = {2003.06137},
 primaryClass = {astro-ph.HE},
       adsurl = {https://ui.adsabs.harvard.edu/abs/2020NewAR..8801539H}
}

@article{WagnerBicknellUmemura2012,
       author = {{Wagner}, A.~Y. and {Bicknell}, G.~V. and {Umemura}, M.},
        title = "{Driving Outflows with Relativistic Jets and the Dependence of Active Galactic Nucleus Feedback Efficiency on Interstellar Medium Inhomogeneity}",
      journal = {\apj},
         year = 2012,
        month = oct,
       volume = {757},
       number = {2},
          eid = {136},
        pages = {136},
          doi = {10.1088/0004-637X/757/2/136},
archivePrefix = {arXiv},
       eprint = {1205.0542},
 primaryClass = {astro-ph.CO},
       adsurl = {https://ui.adsabs.harvard.edu/abs/2012ApJ...757..136W}
}

@article{Cielo2018,
       author = {{Cielo}, Salvatore and {Bieri}, Rebekka and {Volonteri}, Marta and {Wagner}, Alexander Y. and {Dubois}, Yohan},
        title = "{AGN feedback compared: jets versus radiation}",
      journal = {\mnras},
         year = 2018,
        month = jun,
       volume = {477},
       number = {1},
        pages = {1336-1355},
          doi = {10.1093/mnras/sty708},
archivePrefix = {arXiv},
       eprint = {1712.03955},
 primaryClass = {astro-ph.GA},
       adsurl = {https://ui.adsabs.harvard.edu/abs/2018MNRAS.477.1336C}
}

@article{Veilleux2005,
       author = {{Veilleux}, Sylvain and {Cecil}, Gerald and {Bland-Hawthorn}, Joss},
        title = "{Galactic Winds}",
      journal = {\araa},
         year = 2005,
        month = sep,
       volume = {43},
       number = {1},
        pages = {769-826},
          doi = {10.1146/annurev.astro.43.072103.150610},
archivePrefix = {arXiv},
       eprint = {astro-ph/0504435},
 primaryClass = {astro-ph},
       adsurl = {https://ui.adsabs.harvard.edu/abs/2005ARA&A..43..769V}
}

@article{Morganti2017,
       author = {{Morganti}, Raffaella},
        title = "{Radio jets clearing the way through galaxies: the view from Hi and molecular gas}",
    booktitle = {Extragalactic Jets from Every Angle},
         year = 2015,
       editor = {{Massaro}, F. and {Cheung}, C.~C. and {Lopez}, E. and {Siemiginowska}, A.},
       series = {IAU Symposium},
       volume = {313},
        month = mar,
        pages = {283-288},
          doi = {10.1017/S1743921315002331},
archivePrefix = {arXiv},
       eprint = {1411.6107},
 primaryClass = {astro-ph.GA},
       adsurl = {https://ui.adsabs.harvard.edu/abs/2015IAUS..313..283M}
}

@article{Boehringer1993,
       author = {{Boehringer}, H. and {Voges}, W. and {Fabian}, A.~C. and {Edge}, A.~C. and {Neumann}, D.~M.},
        title = "{A ROSAT HRI study of the interaction of the X-ray emitting gas and radio lobes of NGC 1275.}",
      journal = {\mnras},
         year = 1993,
        month = oct,
       volume = {264},
        pages = {L25-L28},
          doi = {10.1093/mnras/264.1.L25},
       adsurl = {https://ui.adsabs.harvard.edu/abs/1993MNRAS.264L..25B}
}

@article{Birzan2004,
       author = {{B{\^\i}rzan}, L. and {Rafferty}, D.~A. and {McNamara}, B.~R. and {Wise}, M.~W. and {Nulsen}, P.~E.~J.},
        title = "{A Systematic Study of Radio-induced X-Ray Cavities in Clusters, Groups, and Galaxies}",
      journal = {\apj},
         year = 2004,
        month = jun,
       volume = {607},
       number = {2},
        pages = {800-809},
          doi = {10.1086/383519},
archivePrefix = {arXiv},
       eprint = {astro-ph/0402348},
 primaryClass = {astro-ph},
       adsurl = {https://ui.adsabs.harvard.edu/abs/2004ApJ...607..800B}
}

@article{Rafferty2006,
       author = {{Rafferty}, D.~A. and {McNamara}, B.~R. and {Nulsen}, P.~E.~J. and {Wise}, M.~W.},
        title = "{The Feedback-regulated Growth of Black Holes and Bulges through Gas Accretion and Starbursts in Cluster Central Dominant Galaxies}",
      journal = {\apj},
         year = 2006,
        month = nov,
       volume = {652},
       number = {1},
        pages = {216-231},
          doi = {10.1086/507672},
archivePrefix = {arXiv},
       eprint = {astro-ph/0605323},
 primaryClass = {astro-ph},
       adsurl = {https://ui.adsabs.harvard.edu/abs/2006ApJ...652..216R}
}

@article{RupkeVeilleuxSanders2005,
       author = {{Rupke}, David S. and {Veilleux}, Sylvain and {Sanders}, D.~B.},
        title = "{Outflows in Infrared-Luminous Starbursts at z < 0.5. II. Analysis and Discussion}",
      journal = {\apjs},
         year = 2005,
        month = sep,
       volume = {160},
       number = {1},
        pages = {115-148},
          doi = {10.1086/432889},
archivePrefix = {arXiv},
       eprint = {astro-ph/0506611},
 primaryClass = {astro-ph},
       adsurl = {https://ui.adsabs.harvard.edu/abs/2005ApJS..160..115R}
}

@article{Martin2005,
       author = {{Martin}, Crystal L.},
        title = "{Mapping Large-Scale Gaseous Outflows in Ultraluminous Infrared Galaxies with Keck II ESI Spectra: Spatial Extent of the Outflow}",
      journal = {\apj},
         year = 2006,
        month = aug,
       volume = {647},
       number = {1},
        pages = {222-243},
          doi = {10.1086/504886},
archivePrefix = {arXiv},
       eprint = {astro-ph/0604173},
 primaryClass = {astro-ph},
       adsurl = {https://ui.adsabs.harvard.edu/abs/2006ApJ...647..222M}
}

@ARTICLE{Boselli2016,
       author = {{Boselli}, A. and {Cuillandre}, J.~C. and {Fossati}, M. and {Boissier}, S. and {Bomans}, D. and {Consolandi}, G. and {Anselmi}, G. and {Cortese}, L. and {C{\^o}t{\'e}}, P. and {Durrell}, P. and {Ferrarese}, L. and {Fumagalli}, M. and {Gavazzi}, G. and {Gwyn}, S. and {Hensler}, G. and {Sun}, M. and {Toloba}, E.},
        title = "{Spectacular tails of ionized gas in the Virgo cluster galaxy NGC 4569}",
      journal = {\aap},
         year = 2016,
        month = mar,
       volume = {587},
          eid = {A68},
        pages = {A68},
          doi = {10.1051/0004-6361/201527795},
archivePrefix = {arXiv},
       eprint = {1601.04978},
 primaryClass = {astro-ph.GA},
       adsurl = {https://ui.adsabs.harvard.edu/abs/2016A&A...587A..68B}
}

@ARTICLE{Machacek2004,
       author = {{Machacek}, Marie E. and {Jones}, Christine and {Forman}, William R.},
        title = "{Chandra Observations of NGC 4438: An Environmentally Damaged Galaxy in the Virgo Cluster}",
      journal = {\apj},
         year = 2004,
        month = jul,
       volume = {610},
       number = {1},
        pages = {183-200},
          doi = {10.1086/421448},
archivePrefix = {arXiv},
       eprint = {astro-ph/0312322},
 primaryClass = {astro-ph},
       adsurl = {https://ui.adsabs.harvard.edu/abs/2004ApJ...610..183M}
}

@ARTICLE{Lyutikov2006,
       author = {{Lyutikov}, M.},
        title = "{Magnetic draping of merging cores and radio bubbles in clusters of galaxies}",
      journal = {\mnras},
         year = 2006,
        month = nov,
       volume = {373},
       number = {1},
        pages = {73-78},
          doi = {10.1111/j.1365-2966.2006.10835.x},
archivePrefix = {arXiv},
       eprint = {astro-ph/0604178},
 primaryClass = {astro-ph},
       adsurl = {https://ui.adsabs.harvard.edu/abs/2006MNRAS.373...73L}
}

@ARTICLE{DursiPfrommer2008,
       author = {{Dursi}, L.~J. and {Pfrommer}, C.},
        title = "{Draping of Cluster Magnetic Fields over Bullets and Bubbles{\textemdash}Morphology and Dynamic Effects}",
      journal = {\apj},
         year = 2008,
        month = apr,
       volume = {677},
       number = {2},
        pages = {993-1018},
          doi = {10.1086/529371},
archivePrefix = {arXiv},
       eprint = {0711.0213},
 primaryClass = {astro-ph},
       adsurl = {https://ui.adsabs.harvard.edu/abs/2008ApJ...677..993D}
}

@ARTICLE{Hota2007,
       author = {{Hota}, Ananda and {Saikia}, D.~J. and {Irwin}, Judith A.},
        title = "{NGC 4438 and its environment at radio wavelengths}",
      journal = {\mnras},
         year = 2007,
        month = sep,
       volume = {380},
       number = {3},
        pages = {1009-1022},
          doi = {10.1111/j.1365-2966.2007.12114.x},
archivePrefix = {arXiv},
       eprint = {0706.3174},
 primaryClass = {astro-ph},
       adsurl = {https://ui.adsabs.harvard.edu/abs/2007MNRAS.380.1009H}
}

@ARTICLE{Vargas2019,
       author = {{Vargas}, Carlos J. and {Walterbos}, Ren{\'e} A.~M. and {Rand}, Richard J. and {Stil}, Jeroen and {Krause}, Marita and {Li}, Jiang-Tao and {Irwin}, Judith and {Dettmar}, Ralf-J{\"u}rgen},
        title = "{CHANG-ES. XVII. H{\ensuremath{\alpha}} Imaging of Nearby Edge-on Galaxies, New SFRs, and an Extreme Star Formation Region{\textemdash}Data Release 2}",
      journal = {\apj},
         year = 2019,
        month = aug,
       volume = {881},
       number = {1},
          eid = {26},
        pages = {26},
          doi = {10.3847/1538-4357/ab27cb},
archivePrefix = {arXiv},
       eprint = {1906.07763},
 primaryClass = {astro-ph.GA},
       adsurl = {https://ui.adsabs.harvard.edu/abs/2019ApJ...881...26V}
}

@ARTICLE{Irwin12,
       author = {{Irwin}, Judith and {Beck}, Rainer and {Benjamin}, R.~A. and {Dettmar}, Ralf-J{\"u}rgen and {English}, Jayanne and {Heald}, George and {Henriksen}, Richard N. and {Johnson}, Megan and {Krause}, Marita and {Li}, Jiang-Tao and {Miskolczi}, Arpad and {Mora}, Silvia Carolina and {Murphy}, E.~J. and {Oosterloo}, Tom and {Porter}, Troy A. and {Rand}, Richard J. and {Saikia}, D.~J. and {Schmidt}, Philip and {Strong}, A.~W. and {Walterbos}, Rene and {Wang}, Q. Daniel and {Wiegert}, Theresa},
        title = "{Continuum Halos in Nearby Galaxies: An EVLA Survey (CHANG-ES). I. Introduction to the Survey}",
      journal = {\aj},
         year = 2012,
        month = aug,
       volume = {144},
       number = {2},
          eid = {43},
        pages = {43},
          doi = {10.1088/0004-6256/144/2/43},
archivePrefix = {arXiv},
       eprint = {1205.5694},
 primaryClass = {astro-ph.CO},
       adsurl = {https://ui.adsabs.harvard.edu/abs/2012AJ....144...43I}
}

@ARTICLE{Irwin12b,
       author = {{Irwin}, Judith and {Beck}, Rainer and {Benjamin}, R.~A. and {Dettmar}, Ralf-J{\"u}rgen and {English}, Jayanne and {Heald}, George and {Henriksen}, Richard N. and {Johnson}, Megan and {Krause}, Marita and {Li}, Jiang-Tao and {Miskolczi}, Arpad and {Mora}, Silvia Carolina and {Murphy}, E.~J. and {Oosterloo}, Tom and {Porter}, Troy A. and {Rand}, Richard J. and {Saikia}, D.~J. and {Schmidt}, Philip and {Strong}, A.~W. and {Walterbos}, Rene and {Wang}, Q. Daniel and {Wiegert}, Theresa},
        title = "{Continuum Halos in Nearby Galaxies: An EVLA Survey (CHANG-ES). II. First Results on NGC 4631}",
      journal = {\aj},
         year = 2012,
        month = aug,
       volume = {144},
       number = {2},
          eid = {44},
        pages = {44},
          doi = {10.1088/0004-6256/144/2/44},
archivePrefix = {arXiv},
       eprint = {1205.5771},
 primaryClass = {astro-ph.GA},
       adsurl = {https://ui.adsabs.harvard.edu/abs/2012AJ....144...44I}
}

@ARTICLE{Guidetti2011,
       author = {{Guidetti}, D. and {Laing}, R.~A. and {Bridle}, A.~H. and {Parma}, P. and {Gregorini}, L.},
        title = "{Ordered magnetic fields around radio galaxies: evidence for interaction with the environment}",
      journal = {\mnras},
         year = 2011,
        month = jun,
       volume = {413},
       number = {4},
        pages = {2525-2544},
          doi = {10.1111/j.1365-2966.2011.18321.x},
archivePrefix = {arXiv},
       eprint = {1101.1807},
 primaryClass = {astro-ph.CO},
       adsurl = {https://ui.adsabs.harvard.edu/abs/2011MNRAS.413.2525G}
}

@ARTICLE{Guidetti2012,
       author = {{Guidetti}, D. and {Laing}, R.~A. and {Croston}, J.~H. and {Bridle}, A.~H. and {Parma}, P.},
        title = "{The magnetized medium around the radio galaxy B2 0755+37: an interaction with the intragroup gas}",
      journal = {\mnras},
         year = 2012,
        month = jun,
       volume = {423},
       number = {2},
        pages = {1335-1350},
          doi = {10.1111/j.1365-2966.2012.20961.x},
archivePrefix = {arXiv},
       eprint = {1203.4582},
 primaryClass = {astro-ph.CO},
       adsurl = {https://ui.adsabs.harvard.edu/abs/2012MNRAS.423.1335G}
}

@ARTICLE{Adebahr2019,
       author = {{Adebahr}, B. and {Brienza}, M. and {Morganti}, R.},
        title = "{Polarised structures in the radio lobes of B2 0258+35. Evidence of magnetic draping?}",
      journal = {\aap},
         year = 2019,
        month = feb,
       volume = {622},
          eid = {A209},
        pages = {A209},
          doi = {10.1051/0004-6361/201833988},
archivePrefix = {arXiv},
       eprint = {1812.07900},
 primaryClass = {astro-ph.GA},
       adsurl = {https://ui.adsabs.harvard.edu/abs/2019A&A...622A.209A}
}

@ARTICLE{Beck05,
       author = {{Beck}, R. and {Krause}, M.},
        title = "{Revised equipartition and minimum energy formula for magnetic field strength estimates from radio synchrotron observations}",
      journal = {Astronomische Nachrichten},
         year = 2005,
        month = jul,
       volume = {326},
       number = {6},
        pages = {414-427},
          doi = {10.1002/asna.200510366},
archivePrefix = {arXiv},
       eprint = {astro-ph/0507367},
 primaryClass = {astro-ph},
       adsurl = {https://ui.adsabs.harvard.edu/abs/2005AN....326..414B}
}

@ARTICLE{CASA22,
       author = {{CASA Team} and {Bean}, Ben and {Bhatnagar}, Sanjay and {Castro}, Sandra and {Donovan Meyer}, Jennifer and {Emonts}, Bjorn and {Garcia}, Enrique and {Garwood}, Robert and {Golap}, Kumar and {Gonzalez Villalba}, Justo and {Harris}, Pamela and {Hayashi}, Yohei and {Hoskins}, Josh and {Hsieh}, Mingyu and {Jagannathan}, Preshanth and {Kawasaki}, Wataru and {Keimpema}, Aard and {Kettenis}, Mark and {Lopez}, Jorge and {Marvil}, Joshua and {Masters}, Joseph and {McNichols}, Andrew and {Mehringer}, David and {Miel}, Renaud and {Moellenbrock}, George and {Montesino}, Federico and {Nakazato}, Takeshi and {Ott}, Juergen and {Petry}, Dirk and {Pokorny}, Martin and {Raba}, Ryan and {Rau}, Urvashi and {Schiebel}, Darrell and {Schweighart}, Neal and {Sekhar}, Srikrishna and {Shimada}, Kazuhiko and {Small}, Des and {Steeb}, Jan-Willem and {Sugimoto}, Kanako and {Suoranta}, Ville and {Tsutsumi}, Takahiro and {van Bemmel}, Ilse M. and {Verkouter}, Marjolein and {Wells}, Akeem and {Xiong}, Wei and {Szomoru}, Arpad and {Griffith}, Morgan and {Glendenning}, Brian and {Kern}, Jeff},
        title = "{CASA, the Common Astronomy Software Applications for Radio Astronomy}",
      journal = {\pasp},
         year = 2022,
        month = nov,
       volume = {134},
       number = {1041},
          eid = {114501},
        pages = {114501},
          doi = {10.1088/1538-3873/ac9642},
archivePrefix = {arXiv},
       eprint = {2210.02276},
 primaryClass = {astro-ph.IM},
       adsurl = {https://ui.adsabs.harvard.edu/abs/2022PASP..134k4501C}}

@ARTICLE{Wardle74,
       author = {{Wardle}, J.~F.~C. and {Kronberg}, P.~P.},
        title = "{The linear polarization of quasi-stellar radio sources at 3.71 and 11.1 centimeters.}",
      journal = {\apj},
         year = 1974,
        month = jan,
       volume = {194},
          eid = {249},
        pages = {249},
          doi = {10.1086/153240},
       adsurl = {https://ui.adsabs.harvard.edu/abs/1974ApJ...194..249W}
}

@ARTICLE{Li2026_model,
       author = {{Li}, Jiang-Tao},
        title = "{Analytical Framework for Expanding Bubbles in a Hot Circumgalactic Medium}",
      journal = {arXiv e-prints},
         year = 2026,
        month = mar,
          eid = {arXiv:2603.21306},
        pages = {arXiv:2603.21306},
          doi = {10.48550/arXiv.2603.21306},
archivePrefix = {arXiv},
       eprint = {2603.21306},
 primaryClass = {astro-ph.HE},
       adsurl = {https://ui.adsabs.harvard.edu/abs/2026arXiv260321306L}
}

@ARTICLE{Mahajan2019,
       author = {{Mahajan}, Smriti and {Ashby}, M.~L.~N. and {Willner}, S.~P. and {Barmby}, P. and {Fazio}, G.~G. and {Maragkoudakis}, A. and {Raychaudhury}, S. and {Zezas}, A.},
        title = "{The Star Formation Reference Survey - III. A multiwavelength view of star formation in nearby galaxies}",
      journal = {\mnras},
         year = 2019,
        month = jan,
       volume = {482},
       number = {1},
        pages = {560-577},
          doi = {10.1093/mnras/sty2699},
archivePrefix = {arXiv},
       eprint = {1810.01336},
 primaryClass = {astro-ph.GA},
       adsurl = {https://ui.adsabs.harvard.edu/abs/2019MNRAS.482..560M}
}

@ARTICLE{Birzan2012,
       author = {{B{\^\i}rzan}, L. and {Rafferty}, D.~A. and {Nulsen}, P.~E.~J. and {McNamara}, B.~R. and {R{\"o}ttgering}, H.~J.~A. and {Wise}, M.~W. and {Mittal}, R.},
        title = "{The duty cycle of radio-mode feedback in complete samples of clusters}",
      journal = {\mnras},
         year = 2012,
        month = dec,
       volume = {427},
       number = {4},
        pages = {3468-3488},
          doi = {10.1111/j.1365-2966.2012.22083.x},
archivePrefix = {arXiv},
       eprint = {1210.7100},
 primaryClass = {astro-ph.CO},
       adsurl = {https://ui.adsabs.harvard.edu/abs/2012MNRAS.427.3468B}
}

@ARTICLE{Panagoulia2014,
       author = {{Panagoulia}, E.~K. and {Fabian}, A.~C. and {Sanders}, J.~S. and {Hlavacek-Larrondo}, J.},
        title = "{A volume-limited sample of X-ray galaxy groups and clusters - II. X-ray cavity dynamics}",
      journal = {\mnras},
         year = 2014,
        month = oct,
       volume = {444},
       number = {2},
        pages = {1236-1259},
          doi = {10.1093/mnras/stu1499},
archivePrefix = {arXiv},
       eprint = {1407.6614},
 primaryClass = {astro-ph.CO},
       adsurl = {https://ui.adsabs.harvard.edu/abs/2014MNRAS.444.1236P}
}

@ARTICLE{Vollmer2009,
       author = {{Vollmer}, B. and {Soida}, M. and {Chung}, A. and {Chemin}, L. and {Braine}, J. and {Boselli}, A. and {Beck}, R.},
        title = "{Ram pressure stripping of the multiphase ISM in the Virgo cluster spiral galaxy NGC 4438}",
      journal = {\aap},
         year = 2009,
        month = mar,
       volume = {496},
       number = {3},
        pages = {669-675},
          doi = {10.1051/0004-6361/200811140},
archivePrefix = {arXiv},
       eprint = {0901.2770},
 primaryClass = {astro-ph.GA},
       adsurl = {https://ui.adsabs.harvard.edu/abs/2009A&A...496..669V}
}

@ARTICLE{Allen2006,
       author = {{Allen}, S.~W. and {Dunn}, R.~J.~H. and {Fabian}, A.~C. and {Taylor}, G.~B. and {Reynolds}, C.~S.},
        title = "{The relation between accretion rate and jet power in X-ray luminous elliptical galaxies}",
      journal = {\mnras},
         year = 2006,
        month = oct,
       volume = {372},
       number = {1},
        pages = {21-30},
          doi = {10.1111/j.1365-2966.2006.10778.x},
archivePrefix = {arXiv},
       eprint = {astro-ph/0602549},
 primaryClass = {astro-ph},
       adsurl = {https://ui.adsabs.harvard.edu/abs/2006MNRAS.372...21A}
}

@ARTICLE{Li2008,
       author = {{Li}, Jiang-Tao and {Li}, Zhiyuan and {Wang}, Q. Daniel and {Irwin}, Judith A. and {Rossa}, Joern},
        title = "{Chandra observation of the edge-on spiral NGC 5775: probing the hot galactic disc/halo connection}",
      journal = {\mnras},
         year = 2008,
        month = oct,
       volume = {390},
       number = {1},
        pages = {59-70},
          doi = {10.1111/j.1365-2966.2008.13749.x},
archivePrefix = {arXiv},
       eprint = {0807.3587},
 primaryClass = {astro-ph},
       adsurl = {https://ui.adsabs.harvard.edu/abs/2008MNRAS.390...59L}
}

@ARTICLE{Heald2022,
       author = {{Heald}, G.~H. and {Heesen}, V. and {Sridhar}, S.~S. and {Beck}, R. and {Bomans}, D.~J. and {Br{\"u}ggen}, M. and {Chy{\.z}y}, K.~T. and {Damas-Segovia}, A. and {Dettmar}, R.-J. and {English}, J. and {Henriksen}, R. and {Ideguchi}, S. and {Irwin}, J. and {Krause}, M. and {Li}, J.-T. and {Murphy}, E.~J. and {Nikiel-Wroczy{\'n}ski}, B. and {Piotrowska}, J. and {Rand}, R.~J. and {Shimwell}, T. and {Stein}, Y. and {Vargas}, C.~J. and {Wang}, Q.~D. and {van Weeren}, R.~J. and {Wiegert}, T.},
        title = "{CHANG-ES XXIII: influence of a galactic wind in NGC 5775}",
      journal = {\mnras},
         year = 2022,
        month = jan,
       volume = {509},
       number = {1},
        pages = {658-684},
          doi = {10.1093/mnras/stab2804},
archivePrefix = {arXiv},
       eprint = {2109.12267},
 primaryClass = {astro-ph.GA},
       adsurl = {https://ui.adsabs.harvard.edu/abs/2022MNRAS.509..658H}
}

@ARTICLE{Xu26,
       author = {{Xu}, Jianghui and {Li}, Jiang-Tao and {Liu}, Guilin and {Luan}, Luan and {Heesen}, Volker and {Beck}, Rainer and {Irwin}, Judith and {Wang}, Q. Daniel and {Stein}, Michael and {Lu}, Li-Yuan and {Yang}, Yang and {Stil}, Jeroen and {English}, Jayanne and {Dettmar}, Ralf-J{\"u}rgen},
        title = "{CHANG-ES. XXXVIII. A Thin Radio Halo Shaped by Slow Cosmic-Ray Transport in the Quiescent Galaxy NGC 4565}",
      journal = {\apj},
         year = 2026,
        month = apr,
       volume = {1001},
       number = {2},
          eid = {164},
        pages = {164},
          doi = {10.3847/1538-4357/ae5636},
archivePrefix = {arXiv},
       eprint = {2603.22417},
 primaryClass = {astro-ph.GA},
       adsurl = {https://ui.adsabs.harvard.edu/abs/2026ApJ..1001..164X}
}

@ARTICLE{Condon92,
       author = {{Condon}, J.~J.},
        title = "{Radio emission from normal galaxies.}",
      journal = {\araa},
         year = 1992,
        month = jan,
       volume = {30},
        pages = {575-611},
          doi = {10.1146/annurev.aa.30.090192.003043},
       adsurl = {https://ui.adsabs.harvard.edu/abs/1992ARA&A..30..575C}
}

@ARTICLE{Ekholm2000,
       author = {{Ekholm}, T. and {Lanoix}, P. and {Teerikorpi}, P. and {Fouqu{\'e}}, P. and {Paturel}, G.},
        title = "{Investigations of the Local Supercluster velocity field. III. Tracing the backside infall with distance moduli from the direct Tully-Fisher relation}",
      journal = {\aap},
         year = 2000,
        month = mar,
       volume = {355},
        pages = {835-847},
          doi = {10.48550/arXiv.astro-ph/0001475},
archivePrefix = {arXiv},
       eprint = {astro-ph/0001475},
 primaryClass = {astro-ph},
       adsurl = {https://ui.adsabs.harvard.edu/abs/2000A&A...355..835E}
}

\end{document}